\documentclass[prd,aps,showpacs,floatfix]{revtex4}

\usepackage{epsfig}
\usepackage{subfigure}

\def\bea{\begin{eqnarray}}
\def\eea{\end{eqnarray}}
\def\be{\begin{equation}}
\def\ee{\end{equation}}

\begin{document}
\voffset 1.5cm
\title{
Neutrino Oscillations and Collider Test of
the R-parity Violating Minimal Supergravity Model
}

\author{
Dong-Won Jung$^1$, Sin Kyu Kang$^2$, Jong Dae Park$^3$ and Eung Jin
Chun$^4$ }

\affiliation{ \\
$^1$ Department of Physics, Korea Advanced Institute of Science and
Technology,
373-1 Kusong-dong, Yusong-gu, Daejoen 305-701, Korea \\
$^2$School of Physics, Seoul Nat'l University, Seoul 151-747, Korea \\
$^3$Institute of Physics and Applied Physics, Yonsei University,
Seoul 120-749, Korea \\
$^4$Korea Institute for Advanced Study, 207-43 Cheongryangri-dong,
Dongdaemun-gu,  Seoul 130-722, Korea}

\begin{abstract}
We study the R-parity violating minimal supergravity models accounting
for the observed neutrino masses and mixing, which can be tested in
future collider experiments. The bi-large mixing can be explained
by allowing five dominant tri-linear couplings $ \lambda'_{1,2,3}$
and $\lambda_{1,2}$.  The desired ratio of the atmospheric and
solar neutrino mass-squared differences can be obtained
in a very limited parameter space where the tree-level contribution
is tuned to be suppressed.  In this allowed region, we quantify
the correlation between the three neutrino mixing angles and
the tri-linear R-parity violating couplings.  Qualitatively,
the relations $| \lambda'_1 | < | \lambda'_2|  \sim | \lambda'_3|$, and
$|\lambda_1| \sim |\lambda_2|$ are required by
the large atmospheric neutrino mixing angle $\theta_{23}$ and
the small angle  $\theta_{13}$, and  the large solar
neutrino mixing angle $\theta_{12}$, respectively.
Such a prediction on the couplings can be tested in the next linear
colliders by observing the branching ratios of the lightest supersymmetric
particle (LSP).
For the stau or the neutralino LSP, the ratio
$|\lambda_1|^2: |\lambda_2|^2: |\lambda_1|^2 + |\lambda_2|^2$
can be measured by establishing
$Br(e\nu): Br(\mu\nu) : Br(\tau\nu)$ or
$Br(\nu e^\pm \tau^\mp ): Br(\nu\mu^\pm\tau^\mp) : Br(\nu\tau^\pm\tau^\mp)$,
respectively.
The information on the couplings $\lambda'_i$ can be
drawn by measuring $Br(l_i t \bar{b}) \propto |\lambda'_i|^2$ if
the neutralino LSP is heavier than the top quark.

\end{abstract}
\pacs{12.60.-i, 14.60.St}
\maketitle

\section{Introduction}
The recent progress in neutrino experiments has established
the picture for three active neutrino oscillations
\cite{Fukuda:1998mi}, which requires new physics beyond the
Standard Model.  The supsersymmetric standard model (SSM)
would be the best motivated candidate for new physics.
In the SSM, the lepton and baryon number conservation is not guaranteed by
the gauge symmetry and thus one usually imposes an extra global symmetry,
the R-parity, to forbid the fast proton decay.   However, the proton longevity
can be ensured by imposing only the baryon number symmetry.  Then, the
neutrino masses and mixing could be the consequence of the lepton
number violation in the SSM.   In this respect, the R-parity violating
SSM has been extensively examined \cite{oldies}.

The phenomenological study of the SSM strongly depends on
the mechanism of supersymmetry (SUSY) breaking.
The most popular scenario would have been the minimal supergravity
model (mSUGRA)
in which the universality of the soft supersymmetry breaking terms
is  introduced to solve the supersymmetric flavor problem.
The mSUGRA model is highly
predictive as it contains only 3 independent soft parameters:
a universal gaugino mass $M_{1/2}$, a universal scalar mass $m_0$,
a universal trilinear coupling $A_0$ at the ultraviolet scale.
The R-parity conserving mSUGRA model has been one of the most popular
model as it contains a natural dark matter candidate of the Universe,
the lightest supersymmetry particle (LSP), on top of its simplicity and
predictability.
Allowing R-parity violation in the mSUGRA model, such a feature is lost
but the LSP can produce clean lepton number  violating signals through
its decay \cite{sig91}, which provides a definite collider test of the model
as the origin of the neutrino masses and  mixing
\cite{mrv,jaja,valle2,cjkp,diaz,dedes}.

This article aims at an extensive investigation of
the mSUGRA model with R-parity violation.
 We will first analyze the parameter
space of the model in which the phenomenological neutrino mass matrix
consistent with the present neutrino experimental results
can be accommodated.  Concerning this, one may separate two issues:
realizing (i) two large and one small mixing angles, and (ii) a
mild hierarchy between the atmospheric and the solar neutrino mass scales.
The property (i) can be explained simply  by introducing an
appropriate lepton flavor structure of the tri-linear
R-parity violating couplings \cite{CK,cjkp}.
Here, let us remark that the bi-large mixing can also be realized with
allowing  only the bilinear R-parity violating terms if  one
introduces non-universality in the soft terms
\cite{valle1,vempati,cjp}.
It is however non-trivial to achieve (ii) as
the mSUGRA model generically predicts very small
values of $\Delta m^2_{sol}/\Delta m^2_{atm}$ where the
atmospheric mass splitting $\Delta m^2_{atm}$
comes basically  from the tree-level contribution over-dominating the
loop corrections which are supposedly responsible for the solar neutrino mass
splitting $\Delta m^2_{sol}$.
As a consequence, we find a very limited
region of the parameter space consistent with neutrino
experimental results.
This is an undesirable feature of the
mSUGRA scenario for the neutrino masses and mixing compared with
other  scenarios such as gauge-mediated SUSY breaking model
\cite{cjkp}.

Given the allowed region of the (R-parity conserving and violating)
parameter space,  we will then  examine various ways to test the model
in the future colliders through the decays of the LSP,
which can be  either a neutralino or a stau.
As discussed in Refs.~\cite{mrv,jaja,valle2,cjkp,diaz,dedes},
in  models of neutrino masses and mixing with R-parity violation, one has
specific predictions for various branching ratios of the LSP decay
as  the structure of lepton flavor violating couplings is dictated
by the pattern of the neutrino mixing.
For this, it is of course necessary that the LSP
has a short lifetime to produce a bunch of decay signals inside
the detector. Since the total LSP decay rate is proportional to
the (heaviest) neutrino mass, the measurement of the LSP decay
length could also be useful to test the model.  In the parameter space
accommodating the neutrino data, we find that the LSP decay length
is shorter than a few centimeters.  The neutrino mixing angles
are basically determined by the ratios of the introduced
trilinear R-parity violating couplings.  We will discuss how strong
correlations they have and how the ratios of the couplings can be
determined by measuring the LSP branching ratios. As we will see,
the quite reliable information of the neutrino mixing angles
can be drawn from the collider experiments if $\tan\beta$ is not too large.
For large $\tan\beta$, the close correlations between the mixing angles
and the trilinear couplings are lost because of
the large tau Yukawa coupling effect.

 This paper is organized as follows.  In Sec.~2, we summarize how
 neutrino mass matrix can be generated up to 1-loop level in the
 context of mSUGRA with generic R-parity violating interactions.
 In the analysis, we take the basis where the bilinear R-parity term is
rotated away.
In Sec.~3, we make a qualitative analysis to examine the sizes of
various R-parity violating couplings in mSUGRA which are required
to account for the current neutrino oscillation data.
In Sec.~4, we calculate the LSP decay rate and branching
ratios of various modes and find how the model can be tested in
the collider experiments.  We will conclude in Sec.~5. Various
useful formulae for the 1-loop contributions to the neutrino
masses and sneutrino vacuum expectation values will be given in
the appendices.

\section{Neutrino mass matrix and flavor structure of trilinear
couplings in mSUGRA}

Let us begin by writing the superpotential in
the basis where the bilinear term $L_i H_2$ is rotated away :

\begin{equation}
W_0 = \mu H_1 H_2 + h^e_i L_i H_1 E^c_i + h^d_i Q_i H_1 D_i^c
 + h^u_i Q_i H_2 U^c_i ,
\end{equation}

\begin{equation}  \label{WRpV}
 W_1=  \lambda_{i} L_i L_3 E^c_3 + \lambda'_{i} L_i Q_3 D^c_3,
\end{equation}
where  $W_0$ is R-parity conserving part and $W_1$ is R-parity
violating part.
Here, we have taken only 5 trilinear couplings, $\lambda_i$ and $\lambda'_i$,
assuming the usual hierarchy of Yukawa couplings.
Among soft SUSY breaking terms, R-parity
violating bilinear terms are given by
\begin{equation} \label{VRpV}
 V_0 = m^2_{L_i H_1} L_i H_1^\dagger +  B_i L_i H_2  + h.c.,
\end{equation}
where $B_i$ is the dimension-two soft parameter.  We will denote
the Higgs bilinear term as $B H_1 H_2$.  In the mSUGRA model,
the bilinear parameters, $m^2_{L_i H_1}$ and $B_i$ vanishes
at the supersymmetry breaking mediation scale, and their non-zero
values at the weak scale are generated through renormalization group
evolution (RGE) which will be included in our numerical calculations.
For the consistent calculation of the Higgs and slepton potential,
we need to include the 1-loop contributions \cite{CK,valle1}
to the scalar potential as follows:
\begin{eqnarray}
V_{1}=\frac{1}{64\pi^2} {\bf Str} {\cal M}^4 \left( \ln\frac{{\cal
M}^2} {Q^2}-\frac{3}{2}\right).
\end{eqnarray}
As is well-known,
the electroweak symmetry breaking gives rise to a
nontrivial vacuum expectation values of sneutrino (SVEV)
$\tilde{\nu_i}$ as follows:
\begin{equation}\label{svev}
 \xi_i \equiv \frac{\langle \tilde{\nu_i} \rangle}
  {\langle H_1^0 \rangle} = - {m^2_{L_i H_1} + B_i t_\beta +
\Sigma_{L_i}^{(1)}
          \over m^2_{\tilde{\nu}_i} + \Sigma_{L_i}^{(2)} },
\end{equation}
where the 1-loop contributions $\Sigma_{L_i}^{(1,2)}$ are given by
$\Sigma_{L_i}^{(1)}=\partial V_1/H_1^{\ast}\partial L_i$,
$\Sigma_{L_i}^{(2)}=\partial V_1/L_i^{\ast}\partial L_i$.
Their explicit forms are presented in appendix B
correcting various mistakes in the original calculation
of Ref.~\cite{CK}.
The bilinear  R-parity violating parameters induce the mixing
between the ordinary particles and superparticles, namely,
neutrinos/neutralinos,
charged leptons/charginos, neutral Higgs bosons/sneutrinos, as
well as charged Higgs bosons/charged sleptons.
The mixing between neutrinos and neutralinos particularly
serves as the origin of the tree-level neutrino masses. We note
that the parameters $\xi_i$ should be very small to account for
tiny neutrino masses. While the effect of such small parameters on
the particle and sparticle mass spectra (apart from the neutrino
sector) are negligible, they induce small but important R-parity
violating vertices between the particles and sparticles, which
in particular destabilizes the LSP together with the original
trilinear couplings, $\lambda_i$ and $\lambda'_i$.
The derivation of the induced R-parity
violating vertices has been presented in Ref. \cite{cjkp}.
>From the seesaw formulae associated with the heavy four
neutralinos, we obtain the light tree-level neutrino mass matrix
of the form ;
\begin{equation} \label{Mtree}
 M^{tree}_{ij} = -{M_Z^2 \over F_N} \xi_i \xi_j \cos^2\beta ,
\end{equation}
where $F_N= M_1M_2/M_{\tilde{\gamma}}+ M_Z^2 \cos{2\beta}/\mu$ with
$M_{\tilde{\gamma}} = c_W^2 M_1 + s_W^2 M_2$.
The R-parity violating vertices between particles and sparticles
can give rise to 1-loop neutrino masses. Including all the 1-loop
corrections, the loop mass matrix can be written as \cite{CK}
 \bea \label{neutrinoloop}
M^{loop}_{ij}= 
-\frac{M_Z^2}{F_N} \left( \xi_i \delta_j +\delta_i \xi_j \right)
\cos\beta +\Pi_{ij},
 \eea
where $\Pi _{ij} $ denotes the 1-loop contribution of the neutrino
self energy and
\bea \label{comentary} \delta_i &=& \Pi_{\nu_i
\widetilde{B}^0} \left(
\frac{-M_2\sin^2\theta_W}{M_{\widetilde{\gamma}} M_W \tan\theta_W}
\right) + \Pi_{\nu_i \widetilde{W}_3} \left( \frac{M_1
\cos^2\theta_W}{M_{\widetilde{\gamma}}M_W} \right) \nonumber \\
&&+\Pi_{\nu_i \widetilde{H}_1^0} \left( \frac{\sin\beta}{\mu}
\right) +\Pi_{\nu_i \widetilde{H}_2^0}\left(
\frac{-\cos\beta}{\mu} \right).
 \eea
Analytic
expressions of  the 1-loop contributions $\Pi's$ are collected in
Appendix A.
Based on the neutrino mass matrix presented in the above, we will
discuss whether the above mass matrix $M^\nu$ can account for both
atmospheric and solar neutrino experimental data.

As we will see, the tree level neutrino masses
are much larger than the  1-loop contributions, which makes it
difficult to account for the solar and atmospheric neutrino
oscillations. As a result, the observed ratio
$\Delta m^2_{sol}/\Delta m^2_{amt}$ can be obtained in
the mSUGRA only when
some cancellation between the tree-level contributions to $\xi_i$ occurs
to make the tree-level neutrino masses
comparable to the 1-loop contributions or even smaller than the
1-loop contributions.

\section{Numerical results: Fitting the neutrino data}

Let us now examine how the parameters in mSUGRA can be
constrained by the recent results from neutrino experiments. For
our numerical analysis, we scan the input parameters in the following
ranges;
  \bea
 100 {\rm GeV} \leq &m_0& \leq 1000 {\rm GeV}, \\
 100 {\rm GeV} \leq &M_{1/2}& \leq 1000 {\rm GeV},\nonumber \\
 0 {\rm GeV} \leq &A_0& \leq 700 {\rm GeV},\nonumber \\
 2 \leq &\tan \beta& \leq 40\nonumber .
 \eea
The ranges for $R-$parity  violating parameters we scan are
 \bea
4 \times 10^{-6} \leq & \left| \lambda_1 \right|& \leq 6 \times
10^{-4}, \\
4 \times 10^{-6} \leq & \left| \lambda_2 \right|&\leq 6 \times
10^{-4},\nonumber \\
3 \times 10^{-9} \leq & \left| \lambda_1^{'} \right|& \leq
10^{-4},\nonumber \\
4 \times 10^{-6} \leq & \left| \lambda_2^{'} \right|& \leq
10^{-3},\nonumber \\
4 \times 10^{-6} \leq & \left| \lambda_3^{'} \right|& \leq
10^{-3}.  \nonumber
\eea
We set the signs of $ M_{1/2}$ and $A_0$ arbitrary,
but find that most of the allowed  parameter space  corresponds to
the case that both signs are positive.

\subsection{General behavior of neutrino oscillation parameters}

To understand the general feature of the R-parity violating mSUGRA model,
we first  plot the ratio $\Delta m^2_{sol}/\Delta m^2_{atm}$ for
randomly generated points in the parameter space specified above.
FIG.~1 shows that our model prefers the ratio in  the range,
$10^{-4}-10^{-6}$, which implies the dominance of the tree-level
mass over the loop mass.  Only a small fraction of the points are within
the allowed range denoted by two horizontal lines.    The big dots
in this range are the solution points accounting for the bi-large
mixing of the solar and atmospheric neutrino oscillations, which can
be basically arranged by appropriate choice of the R-parity violating
couplings.   We note that more solutions can be found for
$\tan\beta \lesssim 15$.
In FIG.~2, we plot $\sin^22\theta_{23}$  in terms of
the ratios $|\xi_2/\xi_3|$ and $|\lambda'_2/\lambda'_3|$
showing  the correlations between
the atmospheric neutrino mixing angle and the ratios of
the   R-parity violating parameters.
The figure shows that the parameter $\xi_i$ has much cleaner correlation
than $\lambda'_i$ for the general points confirming the relation,
$\sin^22\theta_{23} =4 |\xi_2|^2 |\xi_3|^2/\sum_i |\xi_i|^2$ \cite{jaja},
which is a consequence of the tree mass  dominance.
An interesting observation is that $\lambda'_i$ also maintains a rather
clean correlation, which implies that $\lambda'_i$ gives the  larger
contribution to $\xi_i$  than $\lambda_i$.  We also confirmed that
the condition, $|\lambda'_i| < |\lambda'_2| \approx |\lambda'_3|$,
is required to arrange the large atmospheric angle $\theta_{23}$ and
the small CHOOZ angle $\theta_{13}$.

\subsection{Behavior of the solution points}

In order to understand how the solution points can be obtained,
we plot in FIG.~3, the quantity $|m_3^{loop}-m_3^{tree}|/m_3^{tree}$
for the loop corrected and tree-level mass eigenvalue $m^{loop}_3$ and
$m^{tree}_3$, respectively. In the general parameter space (a),  the ratio
is shown to be smaller than 0.1.  The situation becomes quite different
for the solution points (b), for which the loop mass is comparable to, or
even larger than, the tree mass.  This implies that there must be
some cancellation in $\xi_i$ to reduce the tree mass.  As a result,
the clean correlation between $\xi_i$ and the angle $\theta_{23}$ is lost,
which makes it difficult to probe neutrino oscillation through the LSP decays,
as will be discussed later.  In fact,  $\lambda'_i$ will have
the better correlation than $\xi_i$ for the solution points.

\subsection{Importance of SVEV one-loop correction and neutral scalar loops}

In FIG.4, we show how sizable the one-loop contributions to the SVEV
are for the solution points.  As can be seen from two figures,
there occur almost order-one changes, which confirms cancellations
in the  tree-level values of $\xi_i$.   This implies that
the loop correction to
the SVEV should be taken properly into account for
the determination of the neutrino oscillation parameters
in mSUGRA models.

It is also worthwhile to note that the
the loop diagrams with virtual neutral scalars and neutralinos
become sizable opposite to the case where the tree mass dominance works
\cite{cjkp}.
In order to see this, we consider the parameters defined by
\begin{eqnarray}
\eta_i\equiv \xi_i-\frac{B_i}{B},
\end{eqnarray}
which are relevant for the neutral scalar loops \cite{cjp}.
In the general parameter space, one finds $\eta_i$ closely aligned
with $\xi_i$ so that the neutral scalar loops do not give important
contributions. In FIG.~5,
we plot $|(\xi_3-\eta_3)/\eta_3|$ vs.
$|(\xi_2-\eta_2)/\eta_2|$ for the solution points.
The results indicate that $\eta_i$ can be much larger than $\xi_i$ and more
importantly the alignment between $\xi_i$ and $\eta_i$ is lost
in generic points (which are away from the diagonal line).
All of these observations make it complicated
to understand the structure of the neutrino mass matrix from mSUGRA models.
Namely, all the loop corrections have to be properly included in
the determination of the neutrino oscillation parameters.

\subsection{Fit to neutrino data}

Now we are ready to present how the trilinear R-parity violating
couplings are constrained by the mixing angles and
mass-squared differences observed in the
atmospheric and solar neutrino oscillations.
In FIG.~6 and FIG.~7, we show the correlations between the atmospheric
neutrino mixing angle and the ratios $\xi_2/\xi_3$ and
$\lambda'_2/\lambda'_3$, respectively.  As indicated before,
we find a better correlation with the latter.
Even though $\lambda'_2/\lambda'_3$ has no
analytic relation with $\theta_{23}$ for the solution points,
we can obtain the following favorable  ranges through
the parameter scan:
$\left|\lambda'_2/\lambda'_3 \right|$ from the neutrino data
\begin{eqnarray}
 0.4 \lesssim |\lambda'_2/\lambda'_3| \lesssim 2.5 &\mbox{for}&
   \tan\beta = 3-15 \nonumber \\
 0.3 \lesssim |\lambda'_2/\lambda'_3| \lesssim 3.3 & \mbox{for}&
 \tan\beta =  30- 40.
\end{eqnarray}

It is also amusing to find the correlation between
the ratio $\lambda_2/\lambda_3$
and the solar neutrino mixing angle $\theta_{12}$ as shown in FIG.~8.
Similar to the case of the atmospheric neutrino oscillation, we can get
the constraints:
\begin{eqnarray}
  0.3 \lesssim |\lambda_1/\lambda_2| \lesssim 1.6 & \mbox{for} &
\tan\beta=3-15, \nonumber \\
 0.2 \lesssim |\lambda_1/\lambda_2| \lesssim 5.0 & \mbox{for} &
\tan\beta=30-40.
\end{eqnarray}
In addition, fitting the measured mass-squared values,
we find the allowed regions as follows:
\bea
|\lambda_{1,2}|\,, \;\; |\lambda'_{2,3}|
&=& (0.1-2)\times 10^{-4}, \\
|\lambda'_1|  &<& 2.5\times 10^{-5}.
 \eea
Eqs.~(12)-(15) are the key prediction of the mSUGRA model, some of
which can be tested in the future colliders.

\section{Collider test of the model}

In this section, we will investigate how the mSUGRA scenario for
the neutrino masses and mixing can be tested in the collider experiments.
Since the LSP is destabilized by the R-parity violating interactions,
the structure of the R-parity violating couplings drawn in
the above section
may be probed by observing the lepton number violating signals
of the LSP decay.  Based on the parameter sets constrained by the
neutrino data, we first determine the type of the LSP, which can
be either a neutralino or a stau.  Then, we calculate
the cross section for the pair production of the LSP,
and then its decay length and branching ratios.
To consider the observability of the various decay modes, we will
take the luminosity 1000/fb/yr in the next linear colliders
\cite{desy1,desy2}.
For our presentation, we will
show the trilinear R-parity violating parameters and the
resulting neutrino oscillation parameters by selecting
some typical examples of solution points
and then give predictions of the production cross section
and the decay length with important branching ratios of the LSP.
As will be shown later, the LSP production cross sections are of
the order 10-100 fb so that the branching ratios of the order
$10^{-4}-10^{-3}$ will be measurable in the planned linear colliders.

\subsection{Stau LSP}

When the LSP is the stau, $\tilde{\tau}_1$, it mainly decays into two leptons
through the coupling $\lambda_i$.
For small $\tan\beta$, $\tilde{\tau}_1$ is almost the right-handed
stau $\tilde{\tau}_R$  due to the small left-right mixing mass.
Then the light stau  almost decays into leptons
via $\lambda_i L_i L_3 E^c_3$ terms in the superpotential.
Thus, one can expect that the branching ratios of those decay
channels depend on the parameter $\lambda_i$.
In this case, the following relation holds,
 \bea
 Br(e\nu) : Br(\mu \nu) : Br(\tau \nu)
  \simeq
|\lambda_1 |^2 : |\lambda_2 |^2:  |\lambda_1 |^2 + |\lambda_2 |^2,
\label{rel1}
 \eea
as can be seen from Table I. Note that the decay length is much
smaller than  micro-meter($\mu m$) so that the stau LSP production
and decay occur instantaneously. The R-parity violating signals in
the linear collider will be
$$e^+ e^- \to \tilde{\tau}_1 \bar{\tilde{\tau}_1}
\to l^+_i l^-_j \nu \bar{\nu}$$ , which are identical to the
Standard Model background,
$$e^+ e^- \to W^+ W^- \to l^+_i l^-_j \nu \bar{\nu} \,.$$
But this is a flavor independent process and can be deduced
from the total number of events to establish the flavor dependent
quantities $Br(l_i \nu)$.  Therefore, the observation of the following
relations:
\begin{eqnarray}
 Br(\tau \nu) &=& Br(e\nu) + Br(\mu\nu), \nonumber\\
 {Br(e\nu) \over Br(\mu\nu)} &\approx& 0.1-2.6.
\end{eqnarray}
will be a strong indication of the R-parity violation
predicting Eqs.~(17) and (14).

The situation is more complicated for large $\tan\beta$.
A characteristic feature of this case
is that there is a sizable fraction
of the stau LSP decay into top and bottom quarks,
if available kinematically, which is a consequence of
a large left-right stau mixing.  However,
the above relation (\ref{rel1}) becomes obscured by
the large tau Yukawa coupling effect.
In TABLE II, we show this behavior for $\tan\beta=38$.
One can see that the ratio of the branching ratios,
$Br(e\nu) : Br(\mu \nu) : Br(\tau \nu) \simeq 11.5 : 25.5 : 47.2 $,
deviates from the $\lambda_i$ coupling ratio,
$ |\lambda_1 |^2 :
|\lambda_2 |^2:  |\lambda_1 |^2 + |\lambda_2 |^2 \simeq 15.6 :
34.4 : 50.0$.     The deviation from the relation (17) comes from
the large mixing between the stau and the charged Higgs.
Another effect would be the charged Higgs contribution to the event:
$$e^+ e^- \to H^+ H^- \to \tau^+ \tau^- \nu \bar{\nu} \,.$$
Even though a clean prediction for the $\tau$ sector is lost,
we are still able to establish the lepton number violating
signals in the first two generations and measure the quantity:
\begin{equation}
 {Br(e\nu) \over Br(\mu\nu)} =
 \left| {\lambda_1 \over \lambda_2} \right|^2
 \approx 0.04 - 25  \,.
\end{equation}

\subsection{Neutralino LSP}

The lepton number violating signatures from the neutralino decay have
been studied extensively as the LSP is a neutralino in the  most
parameter space \cite{mrv,jaja,valle2,cjkp}.
A characteristic feature of the neutralino LSP is that
the vertex for the process $\tilde{\chi}_1^0 \to  l_i W$
is proportional to $\xi_i$ which determines the tree-level neutrino
mass (6).  As a result,
measuring the branching ratios $Br(l_i jj)$ through either on-shell
 or off-shell $W$ bosons will determine the ratio of $|\xi_i|^2$, that is,
\bea Br(ejj)
: Br(\mu jj) : Br( \tau jj) = |\xi_1|^2 : |\xi_2|^2 : |\xi_3|^2.
\eea
If the tree mass dominates over the loop mass, which is usually
the case in the gauge-mediated supersymmetry breaking models \cite{cjkp},
the neutrino mixing angles $\theta_{23}$ and $\theta_{13}$ can be cleanly
measured in colliders \cite{jaja}.
Unfortunately, this is not the case in the mSUGRA model under consideration.
As we discussed in the previous section (See Fig.~3), the tree mass
has to be suppressed and thus the variables $\xi_i$ are not correlated with
the mixing angles $\theta_{23}$ and $\theta_{13}$ in general and
$\lambda'_i$ maintain better correlations (See Fig.~6 and 7).
>From Tables III-VI, one can see that $Br(l_i jj)$ or $Br(l_i W)$ are
proportional to $\xi_i^2$ which, however, lost  the correlation
with the mixing angles.

On the other hand,  similarly to the stau LSP case,
we can
extract the information on $\lambda_i$ from the measurement of
$\nu l_i^\pm \tau^\mp$ branching ratios for small $\tan \beta$
because the following relation,
\bea Br(\nu e^{\pm} \tau^{\mp}) :
Br( \nu \mu^{\pm} \tau^{\mp}) : Br(\nu \tau^{\pm} \tau^{\mp})
 \simeq |\lambda_1 |^2 :
|\lambda_2 |^2:  |\lambda_1 |^2 + |\lambda_2 |^2, \label{ccc}
\eea
holds as shown in Tables III-VI.
Likewise, if we measure the above branching ratios,
the models can be tested by comparing them with the values allowed
by the neutrino experimental data as in Eq.~(18).  Note that
this is the case for small $\tan\beta < 15$.
For large $\tan\beta$, the relation (21) is invalidated
again because of the large tau Yukawa coupling effect.

Another interesting aspect is that the parameters $\lambda'_i$ can be probed
{\em if the neutralino LSP is heavy enough to allow the decay modes,
$l_i t \bar{b}$}.  The main contributions to these final states
come from the couplings $\lambda'_i$  and thus one obtains the
following approximate relation:
\begin{equation}
 Br(e t\bar{b}) : Br(\mu t\bar{b}) : Br(\tau t\bar{b})  \approx
|\lambda'_1|^2 : |\lambda'_2|^2 : |\lambda'_3|^2
\end{equation}
which should be consistent with the predictions (12), (14) and (15).
In Table V, one finds the branching ratios for $l_i t \bar{b}$ large
enough to be measured in the colliders with the integrated luminosity
of $1000/$fb.   The branching ratios get too small if the LSP mass
is more than twice top quark mass as shown in Table VI.  This is because
the $\nu t\bar{t}$ mode occurring through the light Higgs exchange
 becomes dominating.

\section{Conclusion}

We have investigated the neutrino masses and mixing originated from
the R-parity violation  and its possible collider test in the
context of the minimal supergravity model (mSUGRA) where the soft
terms are assumed to be universal at the ultraviolet scale.  In this
scheme, the bi-large mixing accounting for the atmospheric and solar
neutrino data can be obtained by allowing five independent
tri-linear couplings, $\lambda_i$ and $\lambda'_i$. In the mSUGRA
model, the tree-level contribution to the neutrino mass matrix is
too much large compared with the one-loop contribution in the
generic parameter space.  In order to obtain the observed value of $
\Delta m^2_{sol}/\Delta m^2_{atm} $, the parameters needs to be
tuned to arrange some cancellation in the tree-level mass which is
made to be comparable to the loop mass.   As a result, a nice
relation between the R-parity (and lepton number) violating
parameters and the neutrino oscillation parameters is lost. This is
a dis-favorable feature of the mSUGRA model with R-parity violation
as the origin of the observed neutrino mass matrix. However, in  the
allowed region of the parameter space, we still find the reasonable
correlations between the tri-linear R-parity violating couplings and
the neutrino mixing angles. That is, the bi-large mixing of the
atmospheric and solar neutrino oscillation requires $|\lambda'_2|
\sim | \lambda'_3 |$ (12) and $| \lambda_1| \sim |\lambda_2|$ (13),
while the small mixing angle $\theta_{13}$  requires $|\lambda'_1|
\ll |\lambda'_{2,3}|$ (15).

Such correlations predict some distinctive lepton flavor violating
signatures of the LSP decay in the next linear colliders,
which can provide a test for the R-parity violation scenario.
If the  stau is the LSP,  there is no way to probe the couplings $\lambda'_i$
and thus the two mixing angles $\theta_{23}$ and $\theta_{13}$.
However, it is still possible to obtain the information on the solar
neutrino mixing angle $\theta_{12}$ since the stau decays into
the final states $l_i \nu$ through the couplings $\lambda_i$.
When $\tan\beta$ is not too large, the relation
$ Br(e\nu) : Br(\mu\nu) : Br(\tau\nu) =
|\lambda_1|^2 : |\lambda_2|^2 : |\lambda_1|^2 + |\lambda_2|^2 $
can be checked to confirm or disprove our scenario.
For large $\tan\beta$, the last relation in the $\tau$ sector is
obscured by the large tau Yukawa coupling effect.  Thus, it will not
be possible to establish the above relation without knowing
other parameters like $\tan\beta$, Higgs masses, etc.
For the case of the neutralino LSP,  its decay into $\nu l_i^\pm \tau^\mp$
can be observed to establish the similar relation as above:
 $Br(\nu e^\pm \tau^\mp) : Br(\nu\mu^\pm\tau^\mp) : Br(\nu\tau^\pm\tau^\mp)
= |\lambda_1|^2 : |\lambda_2|^2 : |\lambda_1|^2 + |\lambda_2|^2$.
Again, such a clean relation is invalidated in the $\tau$ sector
for large $\tan\beta$.   If the neutralino is heavier the top quark,
it can decay into $l_i t \bar{b}$ through the couplings $\lambda'_i$,
predicting
$Br(e t \bar{b}) : Br(\mu t \bar{b}) : Br(\tau t \bar{b}) \approx
|\lambda'_1|^2: |\lambda'_2|^2: |\lambda'_3|^2$.
Thus, its measurement can provide information on the
atmospheric neutrino mixing angle.  Let us note that such simple
correlations are valid only for small $\tan\beta \lesssim 15$.

\acknowledgments{ This work was supported by the Korea Research
Foundation Grant, KRF-2002-015-CP0060, EJC was supported by
KRF-2002-070-C00022, and SKK by BK21 program of the Minstry of
Education.

\appendix
\section{}

\renewcommand{\theequation}{\Alph{section}.\arabic{equation}}
\setcounter{equation}{0}

In the appendix A, we present the explicit forms of 1-loop
contributions $\Pi$'s.
Let us begin by fixing the notations for the diagonalization of the mass
matrices.
The mixing between the standard model
particles and the supersymmetric ones in the R-parity violating model can be
rotated away via the so-called seesaw diagonalization which makes the mass
matrices  block-diagonal, and then  each blocks can separately be diagonalized.
\\
\\
\underline{Neutrino-neutralino diagonalization}
\\
\\
The neutrinos/neutralinos mass matrix is written in $7 \times 7$ form as,
\bea
{\cal M}_{\bf N} = \left( \begin{array}{cc} 0 &  M_{\nu N} \\
M_{\nu N}^\dagger &M_{N}
\end{array} \right),
\eea
in the bases
$(\nu_e,\nu_\mu,\nu_\tau,\tilde{B},\tilde{W}_3, \tilde{H}^0_1, \tilde{H}^0_2)$.
 Rotating away the neutrino-neutralino  mixing mass terms
can be made by the following redefinition of neutrinos
and neutralinos with $ \theta^N $
\begin{eqnarray}
\pmatrix{
\nu_i \cr \chi^0_j}  \longrightarrow
\pmatrix{ \nu_i- \theta^N_{ik} \chi^0_k  \cr
\chi^0_j + \theta^N_{lj} \nu_l },
\end{eqnarray}
where $(\nu_i)$ and $(\chi^0_j)$ represent three neutrinos
$(\nu_e, \nu_\mu, \nu_\tau)$ and four neutralinos $(\tilde{B},
\tilde{W}_3, \tilde{H}^0_1, \tilde{H}^0_2)$ in the flavor basis,
respectively.  The rotation parameters $\theta^N_{ij}$ are given by
\begin{eqnarray}
\theta^N_{ij} &=& \xi_i c^N_j c_\beta - \epsilon_i \delta_{j3},
  \nonumber \\
(c^N_j) &=& {M_Z \over F_N} \left({ s_W M_2 \over c_W^2 M_1 + s_W^2 M_2},
-{ c_W M_1 \over c_W^2 M_1 + s_W^2 M_2}, -s_\beta{M_Z\over \mu},
c_\beta{M_Z\over \mu}\right). 
\end{eqnarray}
where  $F_N=M_1 M_2 /( c_W^2 M_1 + s_W^2 M_2) + M_Z^2
s_{2\beta}/\mu$. Here  $s_W=\sin\theta_W$ and $c_W=\cos\theta_W$
with the weak mixing angle $\theta_W$. After rotating away the mixing terms,
the mass matrix has block-diagonal form  with $3\times3$ neutrino mass matrix
 $M_\nu$ and $4\times4$ neutralino mass matrix in each blocks.
In particular, the neutralino mass matrix is diagonalized by $4\times4$ mixing
matrix $N$,
\begin{eqnarray}
N^\dagger M_N N = ({M_N})_{diag}. 
\end{eqnarray}
\\
\\
\underline{Charged lepton/chargino diagonalization}
\\
\\
 SImilar to the above case, charged leptons are also get mixed with charginos.
 Then the $5\times5$ matrix is written as
\bea
{\cal M}_{\bf C} = \left( \begin{array}{cc} M_{ll} & M_{l\tilde{\chi}^\pm} \\
M_{\tilde{\chi}^\pm  l} & M_{\tilde{\chi}^\pm  \tilde{\chi}^\pm}
\end{array} \right),
\eea
in the bases $ \left( e_i,\widetilde{W}^- ,\widetilde{H}_1^- \right)$,
$ \left( e^c_i,\widetilde{W}^+ ,\widetilde{H}_1^+ \right)$.
 Defining  $\theta^L$ and $\theta^R$  as the two rotation matrices
corresponding to the left-handed negatively  and positively
charged fermions, we have
\begin{eqnarray}
\pmatrix{ e_i \cr \chi^-_j } \rightarrow
\pmatrix{ e_i- \theta^L_{ik} \chi^-_k  \cr
\chi^-_j + \theta^L_{lj} e_l } \quad;\quad
\pmatrix{ e^c_i \cr \chi^+_j } \rightarrow
\pmatrix{ e^c_i- \theta^R_{ik} \chi^+_k  \cr
\chi^+_j + \theta^R_{lj} e^c_l },
\end{eqnarray}
where $e_i$ and $e^c_i$ denote the left-handed charged leptons and
anti-leptons, $(\chi^-_j)=(\tilde{W}^-,\tilde{H}^-_1)$ and
$(\chi^+_j)=(\tilde{W}^+,\tilde{H}^+_2)$. The rotation
parameters $\theta^{L,R}_{ij}$ are  given by
\begin{eqnarray}
&& \theta^L_{ij}= \xi_i c^L_j c_\beta-\epsilon_i \delta_{j2}\;,
\quad
\theta^R_{ij}= {m^e_i\over F_C} \xi_i c^R_j c_\beta  \quad\mbox{and} \nonumber\\
&&  (c^L_j)= -{M_W \over F_C} (\sqrt{2}, 2s_\beta{M_W\over
\mu})\;,
\nonumber \\
&&   (c^R_j)= -{M_W  \over F_C} (\sqrt{2}(1-{M_2\over \mu}
t_\beta),
\frac{M_2^2 c^{-1}_\beta}{\mu M_W }+2{M_W \over \mu} c_\beta),
\end{eqnarray}
and $F_C= M_2 + M_W^2 s_{2\beta}/\mu$. After block diagonalizing, the
$2\times 2$ chargino mass matrix $ M_{\tilde{\chi}^\pm  \tilde{\chi}^\pm}$ is diagonalized
\bea
V^\dagger  M_{\tilde{\chi}^\pm  \tilde{\chi}^\pm} U 
= ( M_{\tilde{\chi}^\pm  \tilde{\chi}^\pm})_{diag}.
\eea
\\
\\
\underline{Sneutrino/neutral Higgs boson diagonalization}
\\
\\
Denoting the rotation matrix by $\theta^S$, we get
\begin{eqnarray} \label{thetaS}
\pmatrix{ \tilde{\nu}_i \cr H^0_1 \cr H^0_2 } \rightarrow
\pmatrix{ \tilde{\nu}_i- \theta^S_{i1} H^0_1 -\theta^S_{i2} H^{0*}_2
- \theta^S_{i3} H^{0*}_1 -\theta^S_{i4} H^{0}_2  \cr
H^0_1 + \theta^S_{i1} \tilde{\nu}_i + \theta^S_{i3} \tilde{\nu}^*_i \cr
H^0_2 + \theta^S_{i2} \tilde{\nu}^*_i + \theta^S_{i4} \tilde{\nu}_i
\cr },
\end{eqnarray}
where
\begin{eqnarray}
\theta^S_{i1} &=& -\xi_i - \eta_i s_\beta^2 m_A^2 [ m^4_{\tilde{\nu}_i}
- m^2_{\tilde{\nu}_i}(m_A^2+M_Z^2 s_\beta^2)-m_A^2M_Z^2s_\beta^2c_{2\beta}]
/F_S, \nonumber\\
\theta^S_{i2} &=& + \eta_i s_\beta c_\beta m_A^2 [ m^4_{\tilde{\nu}_i}
- m^2_{\tilde{\nu}_i}(m_A^2+M_Z^2 c_\beta^2)+m_A^2M_Z^2c_\beta^2c_{2\beta}]
/F_S, \nonumber\\
\theta^S_{i3} &=&  -\eta_i s_\beta^2 c_\beta^2 m_A^2 M_Z^2
[ m^2_{\tilde{\nu}_i} - m_A^2 c_{2\beta}]/F_S, \nonumber\\
\theta^S_{i3} &=&  +\eta_i s_\beta^3 c_\beta m_A^2 M_Z^2
[ m^2_{\tilde{\nu}_i} + m_A^2 c_{2\beta}]/F_S,
\end{eqnarray}
with $F_S= (m^2_{\tilde{\nu}_i}-m_h^2)(m^2_{\tilde{\nu}_i}-m_H^2)
(m^2_{\tilde{\nu}_i}-m_A^2)$ and  $m_A, m_h$ and $m_H$ are the
masses of pseudo-scalar, light and heavy neutral scalar Higgs
bosons, respectively. Note that $m_A^2= -B\mu/c_\beta s_\beta$ in
our convention. For our calculation, we assume that all the
R-parity violating parameters are real and thus all $\theta$'s are
real, too.  We also note that the presence of the scalar fields as
well as their complex conjugates in Eq.~(\ref{thetaS}) is due to
the electroweak symmetry breaking, which is expected to be
suppressed with $M_Z^2$. Sneutrino and Higgs mass matrices
 can then separately be diagonalized.
\\
\\
\underline{Charged slepton/charged Higgs boson diagonalization}
\\
\\
Defining $\theta^C$ as the rotation matrix, we have
\begin{eqnarray}
\pmatrix{ \tilde{e}_i \cr \tilde{e}^{c*}_i \cr H^-_1 \cr H^-_2}
\rightarrow
\pmatrix{ \tilde{e}_i- \theta^C_{i1} H^-_1 -\theta^C_{i2} H^{-}_2  \cr
\tilde{e}^{c*}_i- \theta^C_{i3} H^-_1 -\theta^C_{i4} H^{-}_2  \cr
H^-_1 + \theta^C_{i1} \tilde{e}_i + \theta^C_{i3} \tilde{e}^{c*}_i \cr
H^-_2 + \theta^C_{i2} \tilde{e}_i + \theta^C_{i4} \tilde{e}^{c*}_i \cr },
\end{eqnarray}
where
\begin{eqnarray}
\theta^C_{i1} &=& -\xi_i - \eta_i { s_\beta^2 m_A^2 (m^2_{Ri}-m^2_{H^-})
\over (m^2_{H^-}-m^2_{\tilde{e}_{i1}})
(m^2_{H^-}- m^2_{\tilde{e}_{i2}}) }
-\xi_i {m^e_i \mu m^2_{Di}t_\beta
\over (m^2_{H^-}-m^2_{\tilde{e}_{i1}})
(m^2_{H^-}- m^2_{\tilde{e}_{i2}}) },
\nonumber\\
\theta^C_{i2} &=& - \eta_i { s_\beta c_\beta m_A^2 (m^2_{Ri}-m^2_{H^-})
\over (m^2_{H^-}-m^2_{\tilde{e}_{i1}})
(m^2_{H^-}- m^2_{\tilde{e}_{i2}}) }
-\xi_i {m^e_i \mu m^2_{Di}
\over (m^2_{H^-}-m^2_{\tilde{e}_{i1}})
(m^2_{H^-}- m^2_{\tilde{e}_{i2}}) },
\nonumber\\
\theta^C_{i3} &=& + \eta_i { s_\beta^2 m_A^2 m^2_{Di}
\over (m^2_{H^-}-m^2_{\tilde{e}_{i1}})
(m^2_{H^-}- m^2_{\tilde{e}_{i2}}) }
+\xi_i {m^e_i \mu m^2_{Li} (m^2_{Li}-m^2_{H^-})t_\beta
\over (m^2_{H^-}-m^2_{\tilde{e}_{i1}})
(m^2_{H^-}- m^2_{\tilde{e}_{i2}}) },
\nonumber\\
\theta^C_{i4} &=& + \eta_i { s_\beta c_\beta m_A^2 m^2_{Di}
\over (m^2_{H^-}-m^2_{\tilde{e}_{i1}})
(m^2_{H^-}- m^2_{\tilde{e}_{i2}}) }
+\xi_i {m^e_i \mu (m^2_{Li}-m^2_{H^-})
\over (m^2_{H^-}-m^2_{\tilde{e}_{i1}})
(m^2_{H^-}- m^2_{\tilde{e}_{i2}}) }  \,.
\end{eqnarray}
Here, $m_{H^-}$ stands for the charged-Higgs boson mass, and
$m^2_{Li}$, $m^2_{Ri}$ and $m^2_{Di}$ correspond to the LL, RR and
LR components of the $i$-th charged-slepton mass-squared matrix,
respectively, and $m^2_{\tilde{e}_{i1,i2}}$ are their eigenvalues.
We remark that the appearance of $\xi_i$ in $\theta^S_{i1}$ and
$\theta^C_{i1}$ is due to the fact that we performed the rotation
to remove the Goldstone mode from the redefined sneutrino fields.
\\
\\

With the above notations, let us present the 1-loop contributions $\Pi$'s:

\begin{eqnarray}
\Pi_{ij} = &-& \frac{g^2}{32 \pi^2} (t_W N_{1a}-N_{2a})^2
m_{\chi^0_a} \bf{F^a_{ij}} \nonumber \\
&+& \Big\{~~~\frac{g h_\tau}{16 \pi^2} \delta_{i3} \left[
\theta^R_{3m} U_{1a} V_{ma} m_{\tilde{\chi}^0_a} {\bf K^a_{1j}}
-\theta^L_{34}
m_\tau  {\bf K^\tau_{1j}} \right]  \nonumber \\
&& ~~ +\frac{g \lambda_i}{16 \pi^2} \delta_{j3} \left[
\theta^R_{3m} U_{1a} V_{ma} m_{\tilde{\chi}^-_a} {\bf E0^a_{3}}
-\theta^L_{34}
m_\tau  {\bf E0^\tau_{3}} \right] \nonumber \\
&&~~ + \frac{g\lambda_i}{16 \pi^2} \delta_{j3} \left[
-\theta^R_{3m} U_{1a} V_{ma} m_{\tilde{\chi}^-_a} {\bf E0^a_{i}}
+\theta^L_{34}
m_\tau  {\bf E0^\tau_{i}}\right] \nonumber \\
&&~~ +\frac{h_\tau^2}{16 \pi^2} \delta_{i3} \delta_{j3} \left[
\theta^R_{3m} U_{2a}V_{ma} m_{\tilde{\chi}^-_a} {\bf Q^a_{1i}}
-\theta^L_{35}m_\tau {\bf Q^\tau_{1i}} \right] \nonumber \\
&&~~ +\frac{h_i \lambda_j}{16\pi^2} \delta_{i3} \left[
\theta^R_{3m} U_{2a} V_{ma} m_{\tilde{\chi}^-_a} {\bf H0^a_3}
-m_\tau {\bf Q^\tau_{13}} \right] \nonumber \\
&&~~ - \frac{\lambda_i \lambda_j}{16 \pi^2} m_\tau {\bf H0^\tau_3}
\nonumber \\
&&~~ -\frac{\lambda_i^{\prime} \lambda_j^{\prime}}{16 \pi^2} m_b
{\bf P_{\tilde{b} \tilde{b}^c}} \nonumber \\
&& ~~+( i \longleftrightarrow j ) \hspace{1cm} \Big\},  \\
\Pi_{i\tilde{B}} =&-& \frac{g^2}{32 \pi^2} ( t_W N_{1a}-N_{2a}) [~
t_W N_{3a}m_{\tilde{\chi}^0_a} {\bf G^a_{1i}} - t_W N_{4a}
m_{\tilde{\chi}^0_a} {\bf G^a_{2i}} ~] \nonumber \\
&-& \frac{g^2}{16 \pi^2} \frac{t_W}{\sqrt{2}} [~ U_{1a}V_{2a}
m_{\tilde{\chi}^-_a} {\bf K^a_{2i}} -2\theta^R_{im}
U_{1a}V_{ma}m_{\tilde{\chi}^-_a} {\bf H0^a_i}- \delta_{i3} 2
\theta^L_{34}m_\tau {\bf H0^\tau_3} ~] \nonumber \\
&-& \frac{h_i g}{16 \pi^2} \frac{t_W}{\sqrt{2}}
[~~~U_{2a}V_{2a}m_{\tilde{\chi}^-_a} {\bf Q^a_{2i}} -2
\theta^R_{im}
U_{2a} V_{ma} m_{\tilde{\chi}^-_a} {\bf C0^a_i} \nonumber \\
&&~~~~~~~~~~~~+\theta^R_{im}V_{ma}U_{2a} m_{\tilde{\chi}^-_a} {\bf
J^a} -\delta_{i3} \theta^L_{32} m_\tau {\bf J^\tau}
-\delta_{i3}\delta_{j3}m_\tau {\bf K^\tau_{ij}} ~] \nonumber \\
&-& \frac{\lambda_i g}{16 \pi^2} \frac{t_W}{\sqrt2} [~ 2m_\tau
{\bf C0^\tau_3} -m_\tau {\bf E0^\tau_3} ~] \nonumber \\
&-& \frac{\lambda_i^{\prime}}{16 \pi^2} \frac{t_W}{3 \sqrt2} [~ 2
m_b {\bf P_{\tilde{b}^c \tilde{b}^{c*}}} + m_b {\bf
P_{\tilde{b}\tilde{b}^*}} ~],  \\
\Pi_{i\tilde{W^3}} =&-& \frac{g^2}{32 \pi^2} (t_W
N_{1a}-N_{2a})[~-N_{3a}m_{\tilde{\chi}^0_a} {\bf G^a_{1i}} +
N_{4a}
m_{\tilde{\chi}^0_a}{\bf G^a_{2i}} ~] \nonumber \\
&-& \frac{g^2}{16 \pi^2} \frac{1}{\sqrt2}
U_{1a}V_{2a}m_{\tilde{\chi}^-_a} {\bf K^a_{2i}} \nonumber \\
&+& \frac{h_i g}{16 \pi^2} \frac{1}{\sqrt2} [~~~U_{2a}V_{2a}
m_{\tilde{\chi}^-_a} {\bf Q^a_{2i}} + \theta^R_{im} V_{ma}U_{2a}
m_{\tilde{\chi}^-_a} {\bf J^a}  \nonumber \\
&&~~~~~~~~~~~~~-\delta_{i3} \theta^L_{32} m_\tau {\bf J^\tau}
-\delta_{i3} \delta_{j3} m_\tau
{\bf K^\tau_{1j}}~~] \nonumber \\
&+& \frac{\lambda_i g}{16 \pi^2} \frac{1}{\sqrt2} m_\tau {\bf
E0^\tau_3}
\nonumber \\
&+& \frac{\lambda_i^{\prime} g}{16 \pi^2} \frac{1}{\sqrt2} m_b
{\bf P_{\tilde{b} \tilde{b}^*}},  \\
\Pi_{i \tilde{H_1}} =&-& \frac{g^2}{32 \pi^2}(t_WN_{1a}-N_{2a})^2
m_{\tilde{\chi}^0_a}{\bf G^a_{1i}} \nonumber \\
&-& \frac{g h_i}{16\pi^2}[~ \theta^R_{im} U_{1a} V_{ma}
m_{\tilde{\chi}^-_a} {\bf E0^a_i} -\theta^R_{im} V_{ma}U_{1a}
m_{\tilde{\chi}^-_a} {\bf J^a}  \nonumber \\
&&~~~~~~~~~~~~+ \delta_{i3} \theta^L_{31} m_\tau {\bf J^\tau}
-\delta_{i3}\theta^L_{31} m_\tau {\bf E0^\tau_i} ~]
\nonumber \\
&-& \frac{h_i^2}{16\pi^2} \theta^R_{im}U_{2a}V_{ma}
m_{\tilde{\chi}^-_a } {\bf H0^a_i} \nonumber \\
&+& \frac{h_i h_\tau}{16\pi^2} [~\delta_{i3}\theta^L_{32} m_\tau
{\bf H0^\tau_3} +\delta_{i3} m_\tau {\bf Q^\tau_{1k}} \delta_{k3}
~]
\nonumber \\
&+& \frac{\lambda_i h_\tau}{16\pi^2}  2m_\tau {\bf H0^\tau_3}
\nonumber
\\
&+& \frac{\lambda_i^{\prime} h_b}{16\pi^2} 2m_b {\bf P_{\tilde{b}
\tilde{b}^c}},\\
\Pi_{i \tilde{H_2}} =&& \frac{g^2}{32\pi^2} (t_WN_{1a}-N_{2a})^2
m_{\tilde{\chi}^0_a} {\bf G^a_{2i}} \nonumber \\
&-& \frac{g^2}{16 \pi^2} U_{1a}V_{1a} m_{\tilde{\chi}^-_a} {\bf
K^a_{2i}}
\nonumber \\
&-& \frac{h_i g}{16 \pi^2} U_{2a}V_{1a} m_{\tilde{\chi}^-_a} {\bf
Q^a_{2i}}, 
\end{eqnarray}

where the functions defined above are
\begin{eqnarray}
{ \bf F_{ij}^a } &=& \frac{1}{2} \delta_{ij} \left[ B_0 (
m_{\chi^0_a}^2,m_{\tilde{\nu_i}^R}^2 ) -B_0 (
m_{\chi^0_a}^2,m_{\tilde{\nu_i}^I}^2 ) \right] \nonumber \\
&&+Z_{ji} B_0(m^2_{\tilde{\chi}^0_a},m^2_{\tilde{\nu_i}}) +
Z_{ij}B_0(m^2_{\tilde{\chi}^0_a},m^2_{\tilde{\nu_i}}) \nonumber
\\
&&+\frac{1}{2} \theta_{ih} \theta_{jh}
B_0(m^2_{\tilde{\chi}^0_a},m_h^2 ) +\frac{1}{2} \theta_{iH}
\theta_{jH} B_0(m^2_{\tilde{\chi}^0_a},m_H^2 ) \nonumber \\
&&-\frac{1}{2}
\theta_{iA} \theta_{jA} B_0(m^2_{\tilde{\chi}^0_a},m_A^2 ), \nonumber\\
{\bf G_{1i}^a} &=& \theta^S_{i3}
B_0(m^2_{\tilde{\chi}^0_a},m^2_{\tilde{\nu_i}}) +\frac{1}{2}
s_\alpha \theta_{ih} B_0(m^2_{\tilde{\chi}^0_a},m_h^2 ) \nonumber
\\
&&-\frac{1}{2} c_\alpha \theta_{iH}
B_0(m^2_{\tilde{\chi}^0_a},m_H^2 ) +\frac{1}{2} s_\beta
\theta_{iA} B_0(m^2_{\tilde{\chi}^0_a},m_A^2 ), \nonumber\\
{\bf G_{2i}^a} &=& \theta^S_{i2}
B_0(m^2_{\tilde{\chi}^0_a},m^2_{\tilde{\nu_i}}) -\frac{1}{2}
c_\alpha \theta_{ih} B_0(m^2_{\tilde{\chi}^0_a},m_h^2 ) \nonumber
\\
&&-\frac{1}{2} s_\alpha \theta_{iH}
B_0(m^2_{\tilde{\chi}^0_a},m_H^2 ) +\frac{1}{2} s_\beta
\theta_{iA} B_0(m^2_{\tilde{\chi}^0_a},m_A^2 ), \nonumber\\
{\bf E^a_{ij}} &=& \delta_{ij} \left[ c_i^2
B_0(m^2_{\tilde{\chi}^-_a},m^2_{\tilde{e}_{i1}} ) +s_i^2
B_0(m^2_{\tilde{\chi}^-_a},m^2_{\tilde{e}_{i2}} ) \right]
\nonumber \\
&&+\left( \theta^C_{i1} \theta^C_{j1} s_\beta^2 + \theta^C_{i1}
\theta^C_{j2} s_\beta c_\beta +\theta^C_{i2} \theta^C_{j1} c_\beta
s_\beta + \theta^C_{i2} \theta^C_{j2} c_\beta^2 \right) B_0
(m^2_{\tilde{\chi}^-_a},m^2_{H^-} ), \nonumber\\
{\bf H_{ij}^a } &=& -\delta_{ij} c_i s_i
\left[B_0(m^2_{\tilde{\chi}^-_a},m^2_{\tilde{e}_{i1}} ) -
B_0(m^2_{\tilde{\chi}^-_a},m^2_{\tilde{e}_{i1}} ) \right]
\nonumber \\
&&+\left(\theta^C_{i1} \theta^C_{j3} s_\beta^2 + \theta^C_{i1}
\theta^C_{j4} s_\beta c_\beta +\theta^C_{i2} \theta^C_{j3} c_\beta
s_\beta + \theta^C_{i2} \theta^C_{j4} c_\beta^2 \right) B_0
(m^2_{\tilde{\chi}^-_a},m^2_{H^-} ), \nonumber\\
{\bf K_{1i}^a } &=& \left( \theta^C_{i1} c_i^2 - \theta^C_{i3} c_i
s_i \right) B_0( m^2_{\tilde{\chi}^-_a},m^2_{\tilde{e}_{i1}})
+\left( \theta^C_{i1} s_i^2 + \theta^C_{i3} c_i s_i \right) B_0(
m^2_{\tilde{\chi}^-_a},m^2_{\tilde{e}_{i2}}) \nonumber \\
&&-\left( \theta^C_{i1} s_\beta^2 + \theta^C_{i2} c_\beta s_\beta
\right) B_0( m^2_{\tilde{\chi}^-_a},m^2_{H^-}), \nonumber\\
{\bf K_{2i}^a } &=& \left( \theta^C_{i2} c_i^2 - \theta^C_{i4} c_i
s_i \right) B_0( m^2_{\tilde{\chi}^-_a},m^2_{\tilde{e}_{i1}})
+\left( \theta^C_{i2} s_i^2 + \theta^C_{i4} c_i s_i \right) B_0(
m^2_{\tilde{\chi}^-_a},m^2_{\tilde{e}_{i2}}) \nonumber \\
&&-\left( \theta^C_{i1} s_\beta c_\beta + \theta^C_{i2} c_\beta^2
\right) B_0( m^2_{\tilde{\chi}^-_a},m^2_{H^-}), \nonumber\\
{\bf Q^a_{1i} } &=&\left( -\theta^C_{i1} c_i s_i + \theta^C_{i3}
s_i^2 \right) B_0( m^2_{\tilde{\chi}^-_a},m^2_{\tilde{e}_{i1}})
+\left( \theta^C_{i1} s_i c_i + \theta^C_{i3} c_i^2 \right) B_0(
m^2_{\tilde{\chi}^-_a},m^2_{\tilde{e}_{i2}}) \nonumber \\
&&-\left( \theta^C_{i3} s_\beta^2  + \theta^C_{i4} s_\beta c_\beta
\right) B_0( m^2_{\tilde{\chi}^-_a},m^2_{H^-}), \nonumber\\
{\bf Q^\tau_{1i} } &=& {\bf Q^a_{1i} }
( m^2_{\tilde{\chi}^-_a}\rightarrow m_\tau ), \nonumber\\
{\bf Q^a_{2i} } &=&\left( -\theta^C_{i2} c_i s_i + \theta^C_{i4}
s_i^2 \right) B_0( m^2_{\tilde{\chi}^-_a},m^2_{\tilde{e}_{i1}})
+\left( \theta^C_{i2} s_i c_i + \theta^C_{i4} c_i^2 \right) B_0(
m^2_{\tilde{\chi}^-_a},m^2_{\tilde{e}_{i2}}) \nonumber \\
&&-\left( \theta^C_{i3} s_\beta c_\beta  + \theta^C_{i4} c_\beta^2
\right) B_0( m^2_{\tilde{\chi}^-_a},m^2_{H^-}), \nonumber\\
{\bf P_{\tilde{b} \tilde{b^c}} } &=& -c_b s_b
B_0(m_b^2,m^2_{\tilde{b}_1}) +s_b c_b B_0(m_b^2,m^2_{\tilde{b}_2})
\nonumber\\
{\bf P_{\tilde{b}^c \tilde{b^c}^*} } \hspace{-3mm} &=& ~ s_b^2
B_0(m_b^2,m^2_{\tilde{b}_1}) + c_b^2 B_0(m_b^2,m^2_{\tilde{b}_2})
\nonumber\\
{\bf P_{\tilde{b} \tilde{b}^*} } &=& ~c_b^2
B_0(m_b^2,m^2_{\tilde{b}_1}) +s_b^2 B_0(m_b^2,m^2_{\tilde{b}_2}),
\nonumber\\
{\bf C^a_{ij} } &=& \delta_{ij} \left[ s_i^2 B_0(
m^2_{\tilde{\chi}^-_a},m^2_{\tilde{e}_{i1}}) + c_i^2 B_0(
m^2_{\tilde{\chi}^-_a},m^2_{\tilde{e}_{i2}}) \right] \nonumber \\
&&+\left(\theta^C_{i3} \theta^C_{j3} s_\beta^2 + \theta^C_{i3}
\theta^C_{j4} s_\beta c_\beta +\theta^C_{i4} \theta^C_{j3} c_\beta
s_\beta + \theta^C_{i4} \theta^C_{j4} c_\beta^2 \right) B_0(
m^2_{\tilde{\chi}^-_a},m^2_{H^- }), \nonumber\\
{\bf J^a} &=& s_\beta^2 B_0( m^2_{\tilde{\chi}^-_a},m^2_{H^-})
\nonumber \\
&&+ \theta^C_{i1} \theta^C_{j1} \delta_{ij} \left[c_i^2 B_0(
m^2_{\tilde{\chi}^-_a},m^2_{\tilde{e}_{i1}}) + s_i^2 B_0(
m^2_{\tilde{\chi}^-_a},m^2_{\tilde{e}_{i2}}) \right] \nonumber \\
&&- \theta^C_{i1} \theta^C_{j3} \delta_{ij} c_is_i\left[ B_0(
m^2_{\tilde{\chi}^-_a},m^2_{\tilde{e}_{i1}}) - B_0(
m^2_{\tilde{\chi}^-_a},m^2_{\tilde{e}_{i2}}) \right], \nonumber\\
{\bf B0^a_i} &=&c_i^2 B_0(
m^2_{\tilde{\chi}^-_a},m^2_{\tilde{e}_{i1}}) + s_i^2 B_0(
m^2_{\tilde{\chi}^-_a},m^2_{\tilde{e}_{i2}}) \nonumber\\
{\bf B0^\tau_i} &=& {\bf B0^a_i} ( m^2_{\tilde{\chi}^-_a}
\rightarrow
m_\tau ), \nonumber\\
{\bf H0_{i}^a } &=& -c_i s_i
\left[B_0(m^2_{\tilde{\chi}^-_a},m^2_{\tilde{e}_{i1}} ) -
B_0(m^2_{\tilde{\chi}^-_a},m^2_{\tilde{e}_{i1}} ) \right], \nonumber\\
{\bf H0_{i}^\tau } &=&{\bf H0_{i}^a } ( m^2_{\tilde{\chi}^-_a}
\rightarrow m_\tau ), \nonumber\\
{\bf E0^a_{i}} &=&  \left[ c_i^2
B_0(m^2_{\tilde{\chi}^-_a},m^2_{\tilde{e}_{i1}} ) +s_i^2
B_0(m^2_{\tilde{\chi}^-_a},m^2_{\tilde{e}_{i2}} ) \right], \nonumber\\
{\bf E0_{i}^\tau } &=&{\bf E0_{i}^a } ( m^2_{\tilde{\chi}^-_a}
\rightarrow m_\tau ), 
\end{eqnarray}
where
\begin{eqnarray}
\theta_{ih} &=&  -s_\alpha ( \theta^S_{i1} +\theta^S_{i3} )
+c_\alpha ( \theta^S_{i2} +\theta^S_{i4} ) \nonumber  \\
&=& +\xi_i s_\alpha+ \eta_i s_\beta m_A^2 { m_{\tilde{\nu}_i}^2
c_{\alpha-\beta} - M_Z^2 c_{2\beta} c_{\alpha+\beta} \over
(m_{\tilde{\nu}_i}^2-m_h^2) (m_{\tilde{\nu}_i}^2-m_H^2) },
\nonumber\\
\theta_{iH} &=& c_\alpha ( \theta^S_{i1} + \theta^S_{i3} )
+ s_\alpha ( \theta^S_{i2} + \theta^S_{i4} ) \nonumber \\
&=& -\xi_i c_\alpha+ \eta_i s_\beta m_A^2 { m_{\tilde{\nu}_i}^2
s_{\alpha-\beta} - M_Z^2  c_{2\beta} s_{\alpha+\beta}  \over
(m_{\tilde{\nu}_i}^2-m_h^2) (m_{\tilde{\nu}_i}^2-m_H^2) },
\nonumber\\
\theta_{iA} &=& s_\beta (\theta^S_{i1} -\theta^S_{i3} )
+c_\beta ( -\theta^S_{i2} + \theta^S_{i4} ) \nonumber \\
&=& -i \xi_i s_\beta + i \eta_i s_\beta { m_A^2 \over m_A^2-
m_{\tilde{\nu}_i}^2},
\nonumber\\
Z_{ij} &=& \eta_i\eta_j m_A^4 M_Z^2 c_\beta^2 s_\beta^4 \left[
{m_{\tilde{\nu}_i}^2 \over  F_S^i} + {m_{\tilde{\nu}_j}^2  \over
F_S^j} \right]  .
\end{eqnarray}
Here the angle
 $\alpha$ is the usual diagonalization angle of two CP even Higgs
bosons, and $ F_S^i \equiv (m_{\tilde{\nu}_i}^2-m_A^2)
(m_{\tilde{\nu}_i}^2-m_h^2) (m_{\tilde{\nu}_i}^2-m_H^2)$. Recall
that the angle $\alpha$ is defined by $c_{2\alpha}=c_{2\beta}
(m_A^2-M_Z^2)/(m_h^2-m_H^2)$ and $s_{2\alpha}=s_{2\beta}
(m_A^2+M_Z^2)/(m_h^2-m_H^2)$. $c_i = \cos \theta_i, s_i = \sin \theta_i$ are
the components of the diagonalization matrices of the slepton mass square
matrices. $i= \tilde{e},\tilde{\mu},\tilde{\tau}$, each. $c_b ,s_b $ are the
same ones for sbottom mass square matrices.
$B_0$ is the loop function defined as
$B_0 (x,y) = -\frac{x}{x-y} \ln \frac{x}{y} -\ln \frac{x}{Q^2}+1$.

\section{}
We collect here 1-loop contributions to the sneutrino minimization
condition by calculating all the field-dependent particle masses.
As given in the paragraph, the sneutrino vacuum expectation values
are given,
\begin{eqnarray}\label{sveva}
 \xi_i \equiv \frac{\langle \tilde{\nu_i} \rangle}
{\langle H_1^0 \rangle} = - {m^2_{L_i H_1} + B_i t_\beta +
\Sigma_{L_i}^{(1)}
\over m^2_{\tilde{\nu}_i} + \Sigma_{L_i}^{(2)} },
\end{eqnarray}
 where the 1-loop contributions $\Sigma_{L_i}^{(1,2)}$ are given by
$\Sigma_{L_i}^{(1)}=\partial V_1/H_1^{\ast}\partial L_i$,
$\Sigma_{L_i}^{(2)}=\partial V_1/L_i^{\ast}\partial L_i$.
Here, we will give the explicit forms of these $\Sigma_{L_i}$'s.

\subsection{Top squark contribution}
The contributions from the stop mass eigenstates $\tilde{t}_{1,2}$
as follows:
\begin{eqnarray}
&&\hspace{-1.3cm}\Sigma^{(1)}_{L_i}(\tilde{t}_j) ={6\over 64\pi^2}
\left\{
  {4M^2_{Dt} \over 2m^2_{\tilde{t}_j} - m^2_{\tilde{t}_1}-
m^2_{\tilde{t}_2} }
   {m_t \mu_i \over v^2s_{2\beta}/2} \right\}
   m^2_{\tilde{t}_j}\!\left( \ln {m^2_{\tilde{t}_j} \over Q^2}
-1\right),
              \\
&&\hspace{-1.3cm}\Sigma^{(2)}_{L_i}(\tilde{t}_j)= {6\over 64\pi^2}
\left\{
 {1\over2}{M_Z^2\over v^2} + { M^2_{Lt}-M^2_{Rt} \over
    2m^2_{\tilde{t}_j} - m^2_{\tilde{t}_1}- m^2_{\tilde{t}_2} }
   { 8M^2_W-5M_Z^2 \over 6 v^2 } \right\}
   m^2_{\tilde{t}_j}\!\left( \ln{m^2_{\tilde{t}_j} \over Q^2}
-1\right),
\end{eqnarray}
where $M^2_{Lt}= {\cal M}^2_{\tilde{t},11}$, $M^2_{Dt}={\cal
M}^2_{\tilde{t},12}$ and $M^2_{Rt}={\cal M}^2_{\tilde{t},22}$
ignoring  small R-parity violating contributions.

\subsection{ Bottom squark contribution}
The sbottom contributions are
\begin{eqnarray}
 \hspace{0cm}\Sigma^{(1)}_{L_i}(\tilde{b}_j) = {6\over
64\pi^2} \left\{
  2\lambda'_i{m_b\over v_1} + {4M^2_{Db} \over 2m^2_{\tilde{b}_j} -
  m^2_{\tilde{b}_1}- m^2_{\tilde{b}_2} }{A'_i \over v_1} \right\}
  m^2_{\tilde{b}_j}\!\left( \ln {m^2_{\tilde{b}_j} \over Q^2}
-1\right),  \\
 \hspace{0cm}\Sigma^{(2)}_{L_i}(\tilde{b}_j)=  {6\over
64\pi^2} \left\{
 -{1\over2}{M_Z^2\over v^2} + { M^2_{Lb}-M^2_{Rb} \over
 2m^2_{\tilde{b}_j} - m^2_{\tilde{b}_1}- m^2_{\tilde{b}_2} }
 { -4M^2_W+M_Z^2 \over 6 v^2 }  \right\}
 m^2_{\tilde{t}_j}\!\left( \ln {m^2_{\tilde{t}_j} \over Q^2} -1\right),
\end{eqnarray}
where $M^2_{Lb}$, $M^2_{Rb}$, and $M^2_{Db}$ are defined as in the
stop case. The contributions from the first two squark generations
can be obtained by obvious substitutions.

\subsection{Charged slepton or charged Higgs contributions}
Let us denote the charged slepton and Higgs contributions to
$\Sigma$'s as
\begin{eqnarray}
\Sigma^{(1,2)}_{L_j}(\phi) = {4\over 64\pi^2}
 S^{(1,2)}_j(\phi)
 m^2_\phi\!\left(\ln{m^2_\phi\over Q^2}-1\right) \,,
\end{eqnarray}
where $\phi$ stands for the mass eigenstates $\tilde{e}_{i1},
\tilde{e}_{i2}$ and $H^-$. Here $S^{(1)}_j(\phi) \equiv
 (\partial m^2_\phi/\partial u_j)/2v_1$  and
$S^{(2)}_j(\phi) \equiv (\partial m^2_\phi/\partial u_j)/2u_j.$
 $M^2_{L_i}= {\cal M}^2_{i,11}, M^2_{D_i}={\cal
M}^2_{i,12}$ and $M^2_{R_i}={\cal M}^2_{i,22}$. Where $i$ runs
$\tilde{e}, \tilde{\mu}, \tilde{\tau}$.
ignoring  small R-parity violating contributions.

\subsection{Charged Higgs contributions}
For $i=1,2$,
\begin{eqnarray}
 S^{(1)}_j ( H^- )
 = \frac{s_{2\beta} M_W^2( B_i  - m_{L_i H_1}^2 t_\beta)}{ v^2
(M_{L_i}^2-m_{H^-}^2) }.
 \end{eqnarray}
For $i=3$, \begin{eqnarray}
 S^{(1)}_3 (H^- )  = \frac{( c_\beta B_3 - s_{\beta} m_{L_3
 H_1}^2 ) ((m_{H^-}^2-M_{R_3}^2 )(m_\tau^2-2 M_W^2 )  s_\beta + \mu
 m_\tau M_{D_3}^2 )}{v^2(m_{H^-}^2-m_{\tilde{\tau}_1}^2 )
(m_{H^-}^2-m_{\tilde{\tau}_2}^2 )  }.
 \end{eqnarray}
For $i=1,2$, \begin{eqnarray}
 S^{(2)}_i (H^-) = \frac{(m_{H^-}^2-M_{L_i}^2)(2M_W^2-M_Z^2)c_{2\beta}
+
 2  M_W^4 s_{2\beta}^2 }{2 v^2 (m_{H^-}^2-M_{L_i}^2 ) }.
 \end{eqnarray}
For $i=3$, \begin{eqnarray}
 S^{(2)}_3 (H^- ) && \nonumber \\
 =\{&& ( (m_{H^-}^2-m_{\tilde{\tau}_1}^2 )
(m_{H^-}^2-m_{\tilde{\tau}_2}^2 )
 ( 2M_W^2 -M_Z^2 -m_\tau^2 ) c_{2\beta}  \nonumber \\
  &-&2 ( m_{H^-}^2 - M_{R_3}^2 ) s_{2\beta}^2  ( M_W^4 +M_W^2
 m_\tau^2 +m_\tau^4 )
  -2( m_{H^-}^2 -M_{L_3}^2 ) c_\beta^2 \mu^2 m_\tau^2  \nonumber \\
  &+&4 s_\beta c_\beta^2 \mu m_\tau ( -m_\tau^2+2 M_W^2 ) M_{D_3}^2
 \}~
 /~ 2v^2(m_{H^-}^2-m_{\tilde{\tau}_1}^2 )
(m_{H^-}^2-m_{\tilde{\tau}_2}^2 ).
 \end{eqnarray}

 \subsection{Scalar Lepton contributions}
 \begin{eqnarray}
 S^{(1)}_1 ( \tilde{e}_L )  =
 \frac{  M_W^2 ( t_\beta ( m_{H^-}^2 c_{2\beta} +M_{L_1}^2 ) B_1
 + ( m_{H^-}^2 c_{2\beta} -M_{L_1}^2) m_{L_1 H_1}^2
 )}{ v^2 M_{L_1}^2 ( m_{H^-}^2 -M_{L_1}^2 ) }.
 \end{eqnarray}
 \begin{eqnarray}
S^{(1)}_{2,3} ( \tilde{e}_L ) = 0.
 \end{eqnarray}
 \begin{eqnarray}
S^{(1)}_{i} ( \tilde{e}_R ) = 0, ~~~ ~~i= 1,3
 \end{eqnarray}

 \begin{eqnarray}
 S^{(1)}_2 ( \tilde{\mu}_L )  =
 \frac{  M_W^2 ( t_\beta ( m_{H^-}^2 c_{2\beta} +M_{L_2}^2 ) B_2
 + ( m_{H^-}^2 c_{2\beta} -M_{L_2}^2) m_{L_2 H_1}^2
 )}{ v^2 M_{L_2}^2 ( m_{H^-}^2 -M_{L_2}^2 ) }.
 \end{eqnarray}
 \begin{eqnarray}
S^{(1)}_{1,3} ( \tilde{\mu}_L ) = 0.
 \end{eqnarray}
 \begin{eqnarray}
S^{(1)}_{i} ( \tilde{\mu}_R ) = 0 , ~~~~~i= 1,3
 \end{eqnarray}

\begin{eqnarray}
 S^{(2)}_1 ( \tilde{e}_L ) =
 \frac{(m_{H^-}^2-M_{L_1}^2)(M_Z^2M_{L_1}^2+2M_W^4 )
 -2M_W^4 m_{H^-}^2 s_{2\beta}^2}{2 v^2 M_{L_1}^2(m_{H^-}^2-M_{L_1}^2)
}.
 \end{eqnarray}
 \begin{eqnarray}
S^{(2)}_{2,3} ( \tilde{e}_L ) = \frac{-2 M_W^2+M_Z^2}{2 v^2}.
 \end{eqnarray}
 \begin{eqnarray}
S^{(2)}_{i} ( \tilde{e}_R ) = \frac{M_W^2 - M_Z^2}{v^2} , ~~~ ~~i=
1,3
 \end{eqnarray}
\begin{eqnarray}
 S^{(2)}_2 ( \tilde{\mu}_L ) =
 \frac{(m_{H^-}^2-M_{L_2}^2)(M_Z^2M_{L_2}^2+2M_W^4 )
 -2M_W^4 m_{H^-}^2 s_{2\beta}^2}{2 v^2 M_{L_2}^2(m_{H^-}^2-M_{L_2}^2)
}.
 \end{eqnarray}
 \begin{eqnarray}
S^{(2)}_{1,3} ( \tilde{\mu}_L ) = \frac{-2 M_W^2+M_Z^2}{2 v^2}.
 \end{eqnarray}
 \begin{eqnarray}
S^{(2)}_{i} ( \tilde{\mu}_R ) = \frac{M_W^2 - M_Z^2}{v^2} , ~~~
~~i= 1,3
 \end{eqnarray}

For $i=1,2$
 \begin{eqnarray}
 S^{(1)}_{i} (\tilde{\tau}_{1,2} ) = -\frac{m_\tau}{v} \lambda_i \mp
\frac{ M_{D_3}^2  t_\beta
}{m_{\tilde{\tau}_2}^2-m_{\tilde{\tau}_1}^2}\lambda_i.
 \end{eqnarray}

For $i=3$,
 \begin{eqnarray}
S^{(1)}_{3} (\tilde{\tau}_{1,2} )  =
 -\frac{(A) \cdot  B_3 +(B) \cdot m_{L_3 H_1}^2
 \mp \frac{(C) \cdot B_3 + (D) \cdot m_{L_3
H_1}^2}{m_{\tilde{\tau}_2}^2-m_{\tilde{\tau}_1}^2}}{ 2 v^2 c_\beta
m_{\tilde{\tau}_2}^2 m_{\tilde{\tau}_1}^2 ( m_{H^-}^2
-m_{\tilde{\tau}_1}^2)(m_{H^-}^2-m_{\tilde{\tau}_2}^2)},
 \end{eqnarray}
 where,
 \begin{eqnarray}
 (A)
=&&(m_{H^-}^2-m_{\tilde{\tau}_1}^2)(m_{H^-}^2-m_{\tilde{\tau}_2}^2)(
M_{D_3}^2 \mu m_\tau + M_{R_3}^2 s_\beta  M_W^2 ) \nonumber \\
&+& m_{H^-}^2 (M_{R_3}^2 ( M_{R_3}^2 -m_{H^-}^2) +M_{D_3}^4)(2
M_W^2 -c_\beta m_\tau^2) s_\beta c_\beta  \nonumber \\
&+& M_{D_3}^2 m_{H^-}^2 ( m_{\tilde{\tau}_2}^2+
m_{\tilde{\tau}_1}^2-m_{H^-}^2) \mu m_\tau c_\beta^2, \nonumber\\ 
 (B) =&&(M_{R_3}^2 m_{H^-}^4 -M_{R_3}^4 M_{L_3}^2
+M_{R_3}^2M_{D_3}^4 \nonumber \\&& ~~~-M_{R_3}^4 m_{H^-}^2
+M_{L_3}^2 M_{R_3}^2 m_{H^-}^2 -2
M_{D_3}^4m_{H^-}^2 )c_\beta  M_W^2 \nonumber \\
&-&m_{\tilde{\tau}_2}^2m_{\tilde{\tau}_1}^2(m_{H^-}^2-M_{R_3}^2)c_\beta
 m_\tau^2 +M_{D_3}^2 m_{H^-}^2 (m_{H^-}^2-m_{\tilde{\tau}_2}^2-
m_{\tilde{\tau}_1}^2 )s_\beta
c_\beta \mu m_\tau \nonumber \\
&+&m_{H^-}^2(M_{R_3}^2(M_{R_3}^2-m_{H^-}^2)+M_{D_3}^4)(2M_W^2-m_\tau^2)c_\beta^3,
\nonumber\\ 
 (C)
=&-&(m_{H^-}^2-m_{\tilde{\tau}_1}^2)(m_{H^-}^2-m_{\tilde{\tau}_2}^2)
\nonumber
\\ && \hspace{10mm}\times (
M_{D_3}^2(m_{\tilde{\tau}_1}^2+m_{\tilde{\tau}_2}^2) \mu m_\tau
+(2M_{D_3}^4
+M_{R_3}^2(M_{R_3}^2-M_{L_3}^2))s_\beta M_W^2  ) \nonumber \\
&-&M_{D_3}^2 m_{H^-}^2 ( (m_{\tilde{\tau}_1}^2
+m_{\tilde{\tau}_2}^2)(m_{\tilde{\tau}_1}^2
+m_{\tilde{\tau}_2}^2-m_{H^-}^2)- 2 m_{\tilde{\tau}_1}^2
m_{\tilde{\tau}_2}^2) c_\beta^2 \mu m_\tau \nonumber \\
&+&m_{H^-}^2(M_{R_3}^6 -(M_{L_3}^2+m_{H^-}^2)M_{R_3}^4 +(m_{H^-}^2
M_{L_3}^2+3M_{D_3}^4)M_{R_3}^2 \nonumber \\
&&~~~~+(M_{L_3}^2-2 m_{H^-}^2)M_{D_3}^4)
 (m_\tau^2-2 M_W^2)s_\beta
c_\beta^2 ,\nonumber\\ 
 (D) =&-&m_{\tilde{\tau}_1}^2m_{\tilde{\tau}_2}^2( (
M_{R_3}^2-M_{L_3}^2)(M_{R_3}^2-m_{H^-}^2) + 2 M_{D_3}^4)c_\beta
m_\tau^2 \nonumber \\
&+&(-2 M_{D_3}^8 +(-M_{R_3}^4 + 4 M_{R_3}^2 m_{H^-}^2 -2m_{H^-}^4
+3 M_{L_3}^2 M_{R_3}^2 ) M_{D_3}^4 \nonumber \\
&& \hspace{10mm}+
M_{R_3}^2(M_{L_3}^2+m_{H^-}^2)(M_{R_3}^2-m_{H^-}^2)(M_{R_3}^2-M_{L_3}^2))c_\beta
 M_W^2 \nonumber \\
&+&M_{D_3}^2 m_{H^-}^2 ((m_{\tilde{\tau}_1}^2
+m_{\tilde{\tau}_2}^2)(m_{\tilde{\tau}_1}^2
+m_{\tilde{\tau}_2}^2-m_{H^-}^2)- 2 m_{\tilde{\tau}_1}^2
m_{\tilde{\tau}_2}^2) s_\beta c_\beta \mu m_\tau
\nonumber \\
&+&( ~M_{R_3}^6 -(M_{L_3}^2+m_{H^-}^2)M_{R_3}^4 +(m_{H^-}^2
M_{L_3}^2+3M_{D_3}^4)M_{R_3}^2 \nonumber \\
&&~~~~~~~~~~+(M_{L_3}^2-2 m_{H^-}^2)M_{D_3}^4~)
 (m_\tau^2-M_W^2) c_\beta^3.
 \end{eqnarray}

For $ i=1,2$,

 \begin{eqnarray}
 S^{(2)}_i (\tilde{\tau}_{1,2}) =-\frac{M_Z^2}{4 v^2}
  \Big( 1 \mp \frac{(M_{L_3}^2-M_{R_3}^2)(4 M_W^2-3M_Z^2)}{M_Z^2 (
m_{\tilde{\tau}_2}^2-m_{\tilde{\tau}_1}^2)} \Big).
 \end{eqnarray}

For $i=3$,
  \begin{eqnarray}
 S^{(2)}_{3} (\tilde{\tau}_{1,2} ) =-\frac{ (E)
\mp \frac{(F)}{m_{\tilde{\tau}_2}^2-m_{\tilde{\tau}_1}^2}}{4
v^2m_{\tilde{\tau}_2}^2m_{\tilde{\tau}_1}^2
(m_{H^-}^2-m_{\tilde{\tau}_1}^2)(m_{H^-}^2-m_{\tilde{\tau}_2}^2)
},
 \end{eqnarray}
where,
 \begin{eqnarray}
 (E)
=&-&(m_{H^-}^2-m_{\tilde{\tau}_1}^2)(m_{H^-}^2-m_{\tilde{\tau}_2}^2) \nonumber \\
&\times&(m_{\tilde{\tau}_1}^2 m_{\tilde{\tau}_2}^2(2 M_W^2 -
M_Z^2) +2 M_{L_3}^2 \mu^2 m_\tau^2 +4 M_{D_3}^2 \mu  M_W^2 m_\tau
s_\beta -2 M_{R_3}^2 M_W^4) \nonumber \\
&+&4 M_{D_3}^2m_{H^-}^2 (
m_{\tilde{\tau}_1}^2+m_{\tilde{\tau}_2}^2 -m_{H^-}^2)(\mu m_\tau^3
-2 \mu m_\tau  M_W^2 ) s_\beta c_\beta^2  \nonumber \\
&-&8 m_{H^-}^2(M_{D_3}^4+ M_{R_3}^4 - M_{R_3}^2 m_{H^-}^2)
(c_\beta^2 s_\beta^2 M_W^4-c_\beta^4 m_\tau^4 +c_\beta^4
M_W^2 m_\tau^2 ) \nonumber \\
&-& 2m_{H^-}^2 (M_{D_3}^4+M_{L_3}^4 -M_{L_3}^2m_{H^-}^2 )c_\beta^2
\mu^2 m_\tau^2 \nonumber \\
&+& 4((2m_{H^-}^2-M_{R_3}^2 ) M_{D_3}^4 +M_{R_3}^2 (M_{L_3}^2
+m_{H^-}^2)(M_{R_3}^2 -m_{H^-}^2)) c_\beta^2  M_W^2 m_\tau^2
\nonumber \\
&-& 2m_{\tilde{\tau}_1}^2 m_{\tilde{\tau}_2}^2( M_{R_3}^2
-m_{H^-}^2 )   c_\beta^2  m_\tau^4, \nonumber\\ 
 (F) = &&
(m_{H^-}^2-m_{\tilde{\tau}_1}^2)(m_{H^-}^2 -
m_{\tilde{\tau}_2}^2 ) \nonumber \\
 &&  \times \Big[m_{\tilde{\tau}_1}^2 m_{\tilde{\tau}_2}^2
(M_{L_3}^2-M_{R_3}^2 )(3 M_Z^2
-2M_W^2 ) \nonumber \\
&& \hspace{20mm} - 2( m_{\tilde{\tau}_1}^2 m_{\tilde{\tau}_2}^2-
M_{D_3}^4 - M_{L_3}^4) \mu^2 m_\tau^2
\nonumber \\
&& \hspace{20mm} + 4 M_{D_3}^2(M_{L_3}^2+M_{R_3}^2 )s_\beta  \mu
M_W^2 m_\tau
\nonumber \\
&& \hspace{20mm}-2(m_{\tilde{\tau}_1}^2 m_{\tilde{\tau}_2}^2 -
M_{D_3}^4 - M_{R_3}^4)
M_W^4 ~\Big] \nonumber \\
&+&2((m_{\tilde{\tau}_1}^2
+m_{\tilde{\tau}_2}^2)(m_{\tilde{\tau}_1}^2
+m_{\tilde{\tau}_2}^2-m_{H^-}^2) - 2 m_{\tilde{\tau}_1}^2
m_{\tilde{\tau}_2}^2)\nonumber \\
&&\times \Big[ ( M_{L_3}^2 M_{R_3}^2 -m_{H^-}^4) c_\beta^2
m_\tau^4 -2 M_{D_3}^2 m_{H^-}^2 s_\beta c_\beta^2 \mu
(m_\tau^3-2M_W^2 m_\tau) \Big] \nonumber \\
&+& 2 m_{H^-}^2  (s_{2\beta}^2  M_W^4 + c_\beta^4 (4 M_W^2
m_\tau^2 -m_\tau^4)) \nonumber \\
  && \hspace{-20mm} \times \Big[ M_{R_3}^6 -(M_{L_3}^2 + m_{H^-}^2)
M_{R_3}^4 +( m_{H^-}^2 M_{L_3}^2  + 3M_{D_3}^4) M_{R_3}^2
+M_{D_4}^4(M_{L_3}^2-2 m_{H^-}^2) \Big] \nonumber \\
&+& 2m_{H^-}^2 c_\beta^2 \mu^2 m_\tau^2 \nonumber \\
 && \hspace{-20mm} \times \Big[ M_{L_3}^6-(M_{R_3}^2 + m_{H^-}^2)
M_{L_3}^4 +( m_{H^-}^2 M_{R_3}^2  +3M_{D_3}^4) M_{L_3}^2
+M_{D_4}^4(M_{R_3}^2-2 m_{H^-}^2) \Big] \nonumber \\
&+& \Big[ 8 M_{D_3}^8 +4 ( M_{R_3}^4 -( 3 M_{L_3}^2 +4 m_{H^-}^2)
M_{R_3}^2 + 2 m_{H^-}^4) M_{D_3}^4 \nonumber \\
 && \hspace{10mm} -4M_{R_3}^2(M_{L_3}^2 +
m_{H^-}^2)(M_{R_3}^2-M_{L_3}^2 )( M_{R_3}^2 -m_{H^-}^2) \Big]
c_\beta^2  M_W^2 m_\tau^2.
 \end{eqnarray}
\subsection{Neutrino-Neutralino contributions}
The sneutrino VEVs contributions are
\begin{eqnarray}
\Sigma^{(1,2)}_{L_i}(\psi)= -{8\over 64\pi^2}
  S^{(1,2)}_i(\psi)
 m^2_\psi \!\left(\ln{m^2_\psi\over Q^2}-1\right),
\end{eqnarray}
where $\psi$ runs for four neutralinos $\tilde{\chi}^0_i$.
 For 4 neutralinos, $\Psi = \tilde{\chi}_{i}^0, i=1,4,$

 \begin{eqnarray}
 S^{(2)}_i (\Psi) = \frac{2\mu M_Z^2}{v^2} m_\Psi \Big[
&-& \mu M_{\tilde{\gamma}} m_\Psi^3 +(-s_{2\beta} M_Z^2
M_{\tilde{\gamma}} + \mu(M_1^2 c_W^2 + M_2^2 s_W^2 )) m_\Psi^2
\nonumber \\
&+&M_{\tilde{\gamma}}(s_{2\beta} M_Z^2 M_{\tilde{\gamma}} +\mu (
M_Z^2 +\mu)) m_\Psi \nonumber  \\  &-&( M_Z^2 M_{\tilde{\gamma}}^2
+\mu^2 ( M_1^2 c_W^2 +M_2^2 s_W^2)) ~\Big] ~/~D(\Psi),
 \end{eqnarray}
where,

 \begin{eqnarray} D(\Psi) =  \det(M_N) \Big[ && 4 m_\Psi^3 -3
(M_1+M_2) m_\Psi^2  \nonumber \\ &+& 2 ( M_1 M_2 -M_Z^2 -\mu^2 )
m_\Psi \nonumber \\ &+& M_Z^2 (M_{\tilde{\gamma}} +s_{2\beta} \mu
) + \mu^2 (M_1 +M_2) ~\Big], \end{eqnarray}
 \begin{eqnarray}
 M_{\tilde{\gamma}} = M_1 c_W^2 + M_2 s_W^2,
 \end{eqnarray}
 \begin{eqnarray}
 \det{M_N} = -\mu( \mu M_1 M_2 + s_{2\beta} ( M_1 M_W^2 + M_2 (
M_Z^2-M_W^2 ))).
 \end{eqnarray}

\subsection{ CP-even sneutrino or Higgs boson contributions}
As in the above case, we write
\begin{eqnarray}
\Sigma^{(1,2)}_{L_i}(\phi) = {2\over 64\pi^2}
 S^{(1,2)}_i(\phi)
 m^2_\phi\!\left(\ln{m^2_\phi\over Q^2}-1\right), \,
\end{eqnarray}
where $\phi$ runs for ${\rm Re}(\tilde{\nu})$, $h^0$ and $H^0$ and
$S^{(1),(2)}_i(\phi)$ are calculated as follows:
 \begin{eqnarray}
S^{(1)}_j (\tilde{\nu_i})= -\frac{\delta_{ij}}{F_i}
\frac{M_Z^2}{v^2} \Big[ (m_A^2 c_{2\beta} -M_{\tilde{\nu}_i}^2 )
m_{L_i H_1}^2 + t_\beta (m_A^2 c_{2\beta} +M_{\tilde{\nu}_i}^2 )
B_i \Big],
 \end{eqnarray}

 \begin{eqnarray}
 S^{(2)}_j (\tilde{\nu_i}) = \frac{M_Z^2}{2 v^2} \Big[ 1 +
\delta_{ij} \big( \frac{3 M_{\tilde{\nu}_i}^4 +(3 m_A^2 + M_Z^2 )
M_{\tilde{\nu}_i}^2 - m_{H^0}^2 m_{h^0}^2}{F_i} -1 \big) \Big],
 \end{eqnarray}

 \begin{eqnarray}
S^{(1)}_i (h^0,H^0 ) = \frac{M_Z^2}{v^2} \frac{1}{F_i} \Big[&& (
(m_A^2 c_{2\beta} -M_{\tilde{\nu}_i}^2 ) m_{L_i H_1}^2 + t_\beta
( m_A^2 c_{2\beta} +M_{\tilde{\nu}_i}^2 ) B_i \big) \nonumber \\
 &\mp&  \frac{1}{ m_{H^0}^2-m_{h^0}^2 }  \{     t_\beta ( -(m_{H^0}^2
+m_{h^0}^2 +2 m_A^2 c_{2\beta}
 )M_{\tilde{\nu}_i}^2 \nonumber \\ &&~~~~+ m_A^2 ( m_A^2 c_{2\beta} +
M_Z^2 c_{2\beta}
( 4c_\beta^2-1)) )  B_i  \nonumber \\
 &&~~~~+ (( m_{H^0}^2 +m_{h^0}^2 -2 m_A^2 c_{2\beta} )
M_{\tilde{\nu}_i}^2 \nonumber \\&&~~~~+ m_A^2 c_{2\beta} ( m_A^2
-M_Z^2 ( 4 c_\beta^2 -3))) m_{L_i H_1}^2  \}    \Big],
 \end{eqnarray}

\begin{eqnarray}
 S^{(2)}_i (h^0,H^0) = -\frac{M_Z^2}{2v^2 F_i} \Big[ && M_Z^2
(M_{\tilde{\nu}_i}^2 - M_A^2 c_{2\beta}^2 ) \nonumber \\
&\mp& \frac{1}{m_{H^0}^2 -m_{h^0}^2 } \{~ c_{2\beta} (
m_A^2-M_Z^2) M_{\tilde{\nu}_i}^4  \nonumber \\ &&~~~~~+( -m_A^4
c_{2\beta} +M_Z^2 m_A^2 (2 s_{2\beta}^2-1) + 2M_Z^4 c_\beta^2 )
M_{\tilde{\nu}_i}^2
\nonumber \\
 && ~~~~~\hspace{0mm} + 2M_Z^2 m_A^2 c_\beta c_{2\beta} ( m_A^2-M_Z^2)
~\}~
 \Big],
 \end{eqnarray}
where,
 \begin{eqnarray}
 F_i = M_{\tilde{\nu}_i}^4 -( m_{H^0}^2+ m_{h^-}^2)
M_{\tilde{\nu}_I}^2 + m_{H^0}^2 m_{h^0}^2.
 \end{eqnarray}
 Also note that $M^2_{\tilde{\nu}_i}$'s are the usual $R$-parity
  conserving sneutrino masses.
\subsection{ CP-odd sneutrino or Higgs boson contributions}
\begin{eqnarray}
\Sigma^{(1,2)}_{L_j}(\phi) = {2\over 64\pi^2}
 S^{(1,2)}_i(\phi)
 m^2_\phi\!\left(\ln{m^2_\phi\over Q^2}-1\right), \,
\end{eqnarray}
where $\phi$ runs for ${\rm Im}(\tilde{\nu})$, $A^0$. Here,
$S^{(1),(2)}_i(\phi)$ are given by \begin{eqnarray}
&&S^{(1)}_j(\tilde{\nu}^I_i)= 0, \\
&&S^{(2)}_j(\tilde{\nu}^I_i)=\frac{1}{2}\frac{M_Z^2}{v^2},\\
&&S^{(1)}_i({A^0})=0,\\
&&S^{(2)}_i({A^0})=-\frac{1}{2}\frac{M_Z^2}{v^2}\cos{2\beta}.
\end{eqnarray}

\subsection{ Charged lepton or chargino contribution}
Neglecting the contributions from the two light charged leptons by
taking $m_{e,\mu}=0$, we get
\begin{eqnarray}
\Sigma^{(1,2)}_{L_i}(\Psi)= -{8\over 64\pi^2} S^{(1,2)}_i(\Psi)
 m^2_\Psi\!\left(\ln{m^2_\Psi\over Q^2}-1\right),
\end{eqnarray}
where $\Psi$ runs for $\tau$ and the charginos  $\tilde{\chi}^-_{1,2}$ and
$S^{(1,2)}(\Psi)$ are given by

 \begin{eqnarray} S_i^{(1)} (\Psi) = - \lambda_i m_\tau
\frac{m_{\Psi}^2}{v_1} \Big[
(m_{\Psi}^2-m_{\tilde{\chi}^+_1}^2)(m_{\Psi}^2-m_{\tilde{\chi}^+_2}^2)
\Big]/D(\Psi), \end{eqnarray}

\begin{eqnarray}
 S^{(2)}_i (\Psi) = 2 \frac{m_{\Psi}^2}{v} \Big[&& m_{\Psi}^4 M_W^2
-m_{\Psi}^2M_W^2 ( \mu^2 + 2 M_W^2 s_\beta^2 \nonumber \\&+&
m_\tau^2) + m_\tau^2 M_W^2 (\mu^2 + 2 M_W^2 s_\beta^2 ) \nonumber
\\ &+& \delta_{ij} m_\tau^2 \{~ m_{\Psi}^4/2c_\beta^2 + m_{\Psi}^2
[
M_W^2 (3 -t_\beta^2 ) -M_2^2/c_\beta^2 ] \nonumber \\
&&~~~~~~~~~~~~~~~~-2M_W^2 (M_2 \mu t_\beta -3 M_W^2 s_\beta^2
-\mu^2)\} ~\Big] /D(\Psi),
 \end{eqnarray}

 where,
 \begin{eqnarray}
 D(\Psi) = && 5 m_{\Psi}^6 - 4m_{\Psi}^4 ( m_{\tilde{\chi}^-_1 }^2
+m_{\tilde{\chi}^-_2}^2 +m_\tau^2 )  \nonumber \\&+& 3 m_{\Psi}^2 [ m_\tau
^2 ( m_{\tilde{\chi}^-_1}^2 + m_{\tilde{\chi}^-_2}^2) + m_{\tilde{\chi}^-_1}^2
m_{\tilde{\chi}^-_2}^2 ] -2 m_\tau^2 m_{\tilde{\chi}^-_1}^2m_{\tilde{\chi}^-_2}^2.
 \end{eqnarray}



\newpage

\begin{figure}\label{ratio}
\epsfig{file=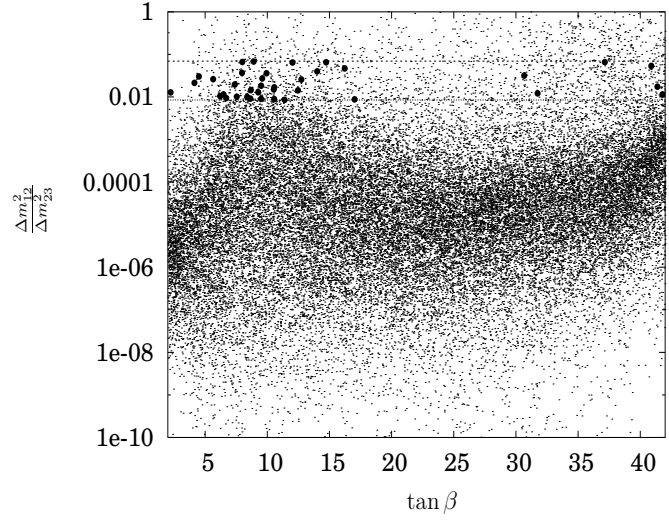,height=7.5cm,width=9.5cm}
\caption{ The ratio $\Delta m^2_{sol}/\Delta m^2_{atm}$ for general
points.  The region between the two straight lines is allowed by the
neutrino data and big dots are the solution points accomodating
the observed mixing angles.}
\end{figure}

\begin{figure}
\subfigure[$\sin^2 2 \theta_{23}$ vs. $ \left| \xi_2/\xi_3 \right|
$]{\epsfig{file=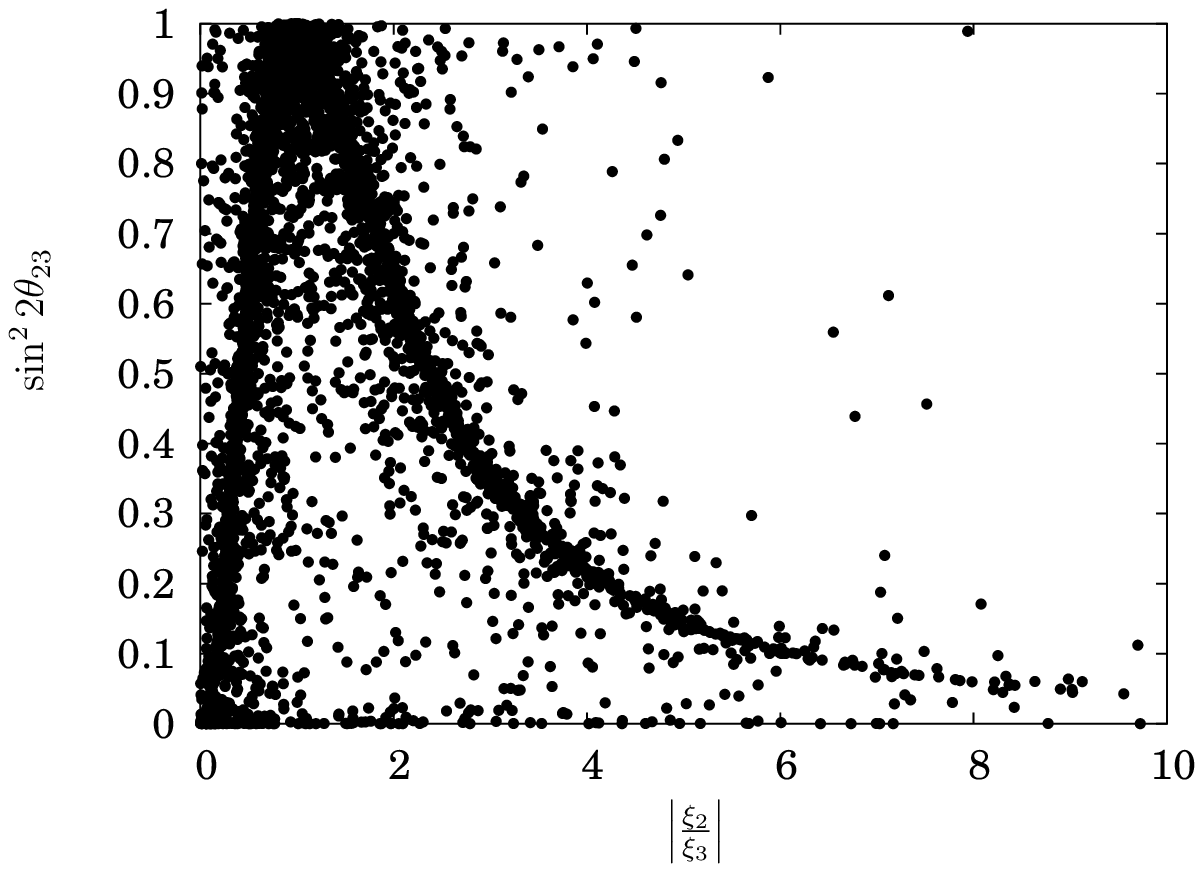 ,height=6.0cm,width=6.0cm}}
\subfigure[$\sin^2 2 \theta_{23}$vs. $ \left| \lambda_2^{'} /
\lambda_3^{'} \right| $ ] {\epsfig{file=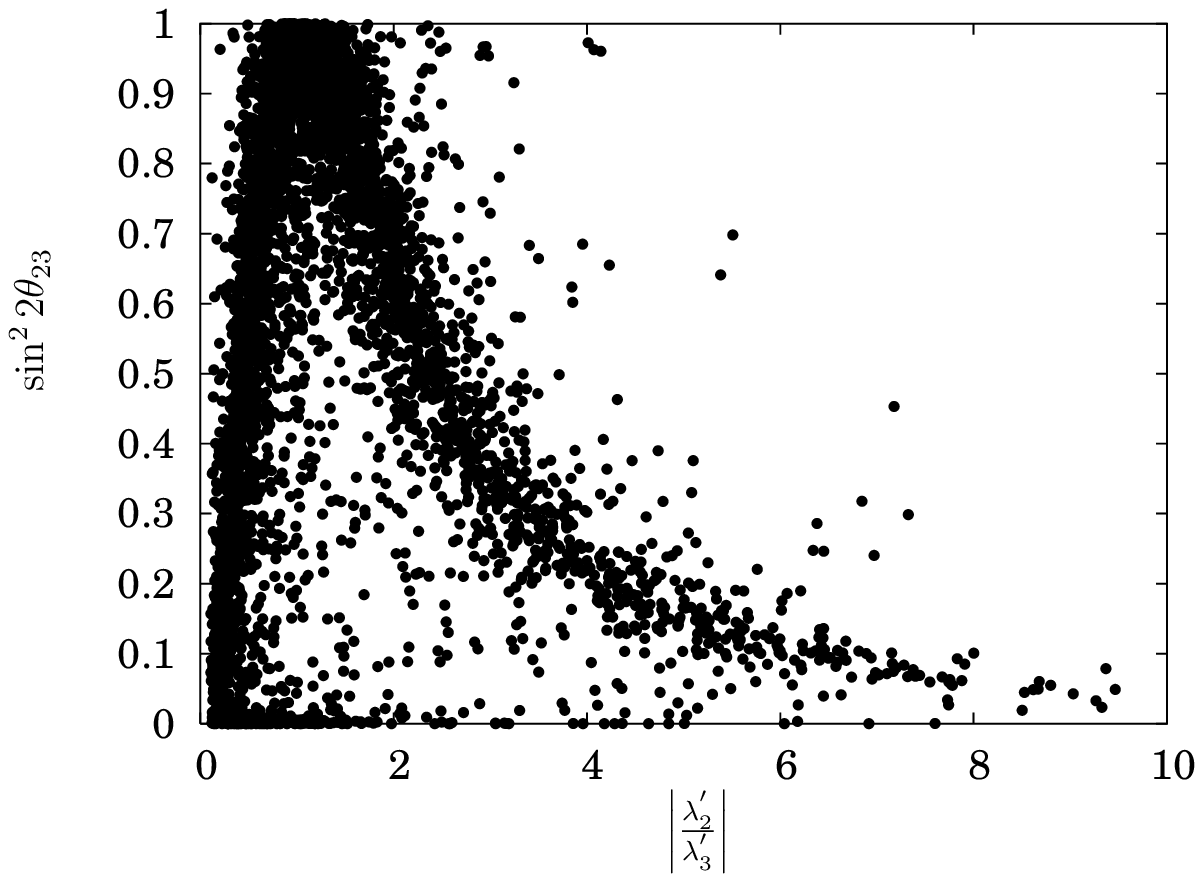
,height=6.0cm,width=6.0cm}}
\caption{ The atmospheric neutrino mixing angle vs. $ \left|
\xi_2/\xi_3 \right| $ and $ | \lambda_2^{'} / \lambda_3^{'}| $
for general points. }
\end{figure}

\begin{figure}
\subfigure[For general points]
{\epsfig{file=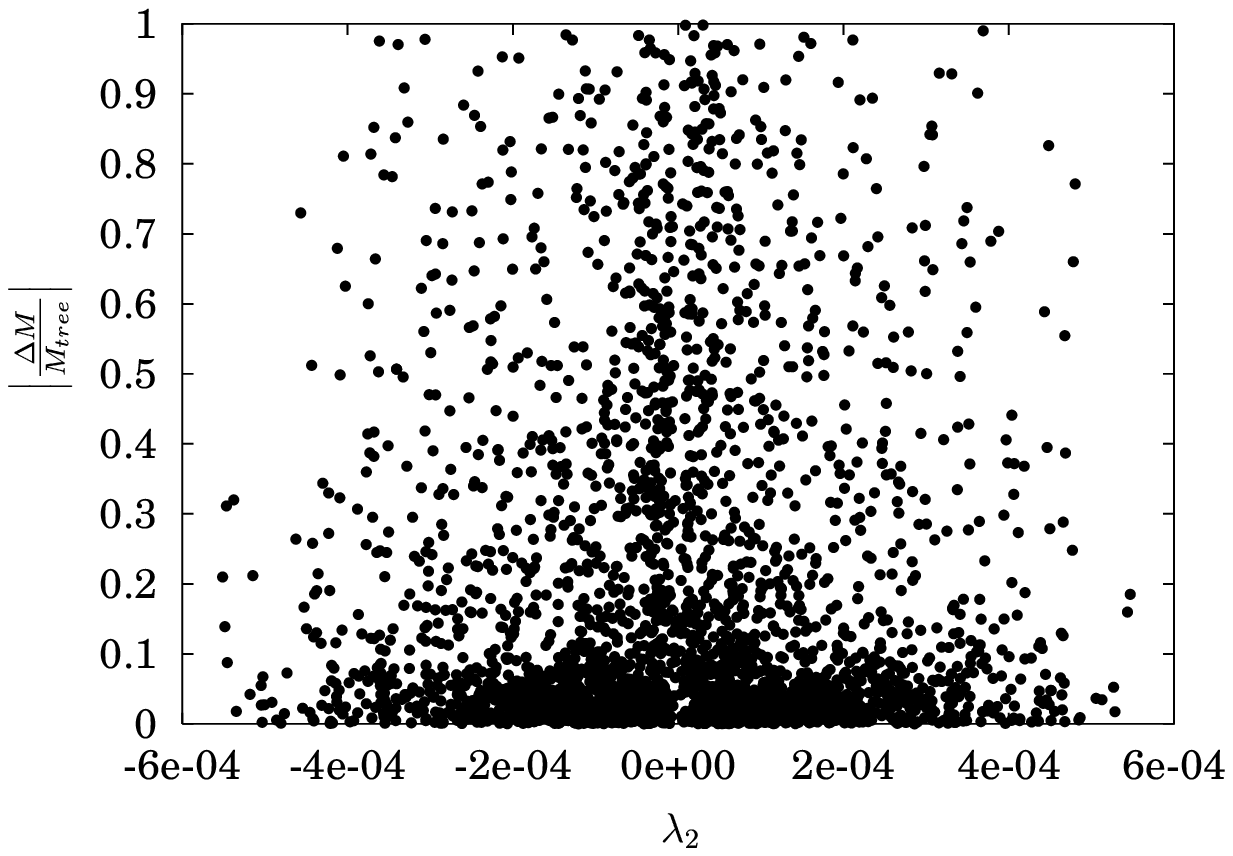, height=6.0cm,width=6.0cm}}
\subfigure[For solution points]
{\epsfig{file=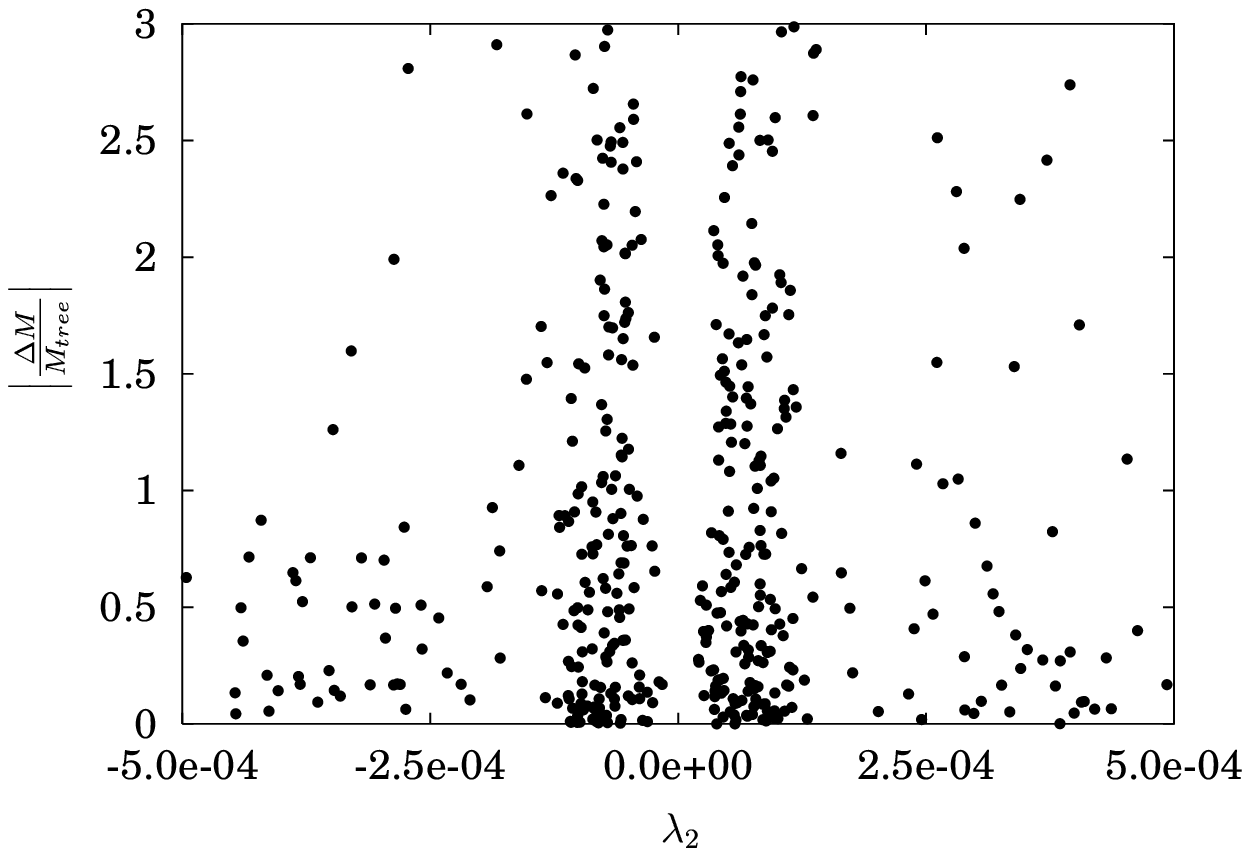, height=6.0cm,width=6.0cm}}
 \caption{ The ratio $|m_3 - m_3^0|/m_3^0$
for the loop-corrected ($m_3$) and
tree-level ($m^0_3$)  neutrino masses
with varying $\lambda_2$. }
  \end{figure}

\begin{figure}
\subfigure[For $\tan \beta = 3 -
15$]{\epsfig{file=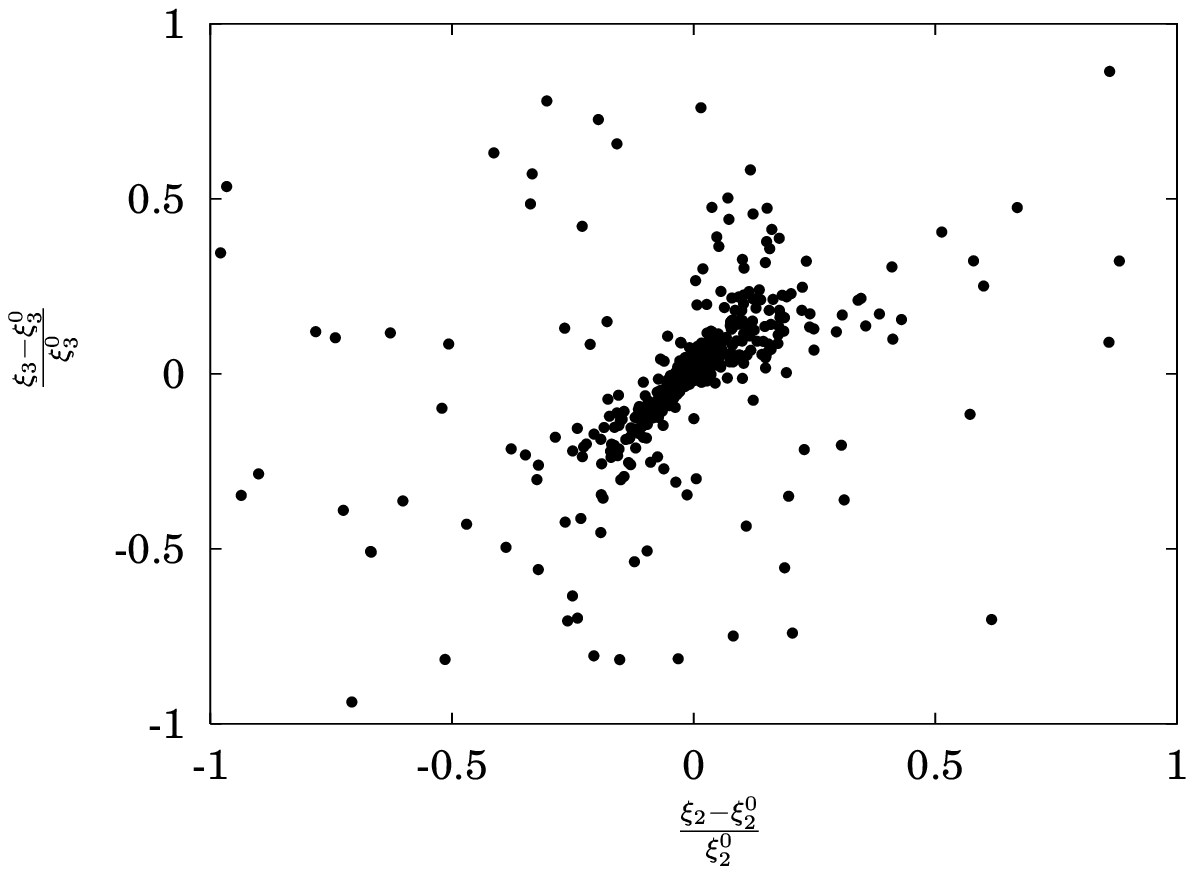 ,height=6.0cm,width=6.0cm}}
\subfigure[For $\tan \beta = 30 -
40$]{\epsfig{file=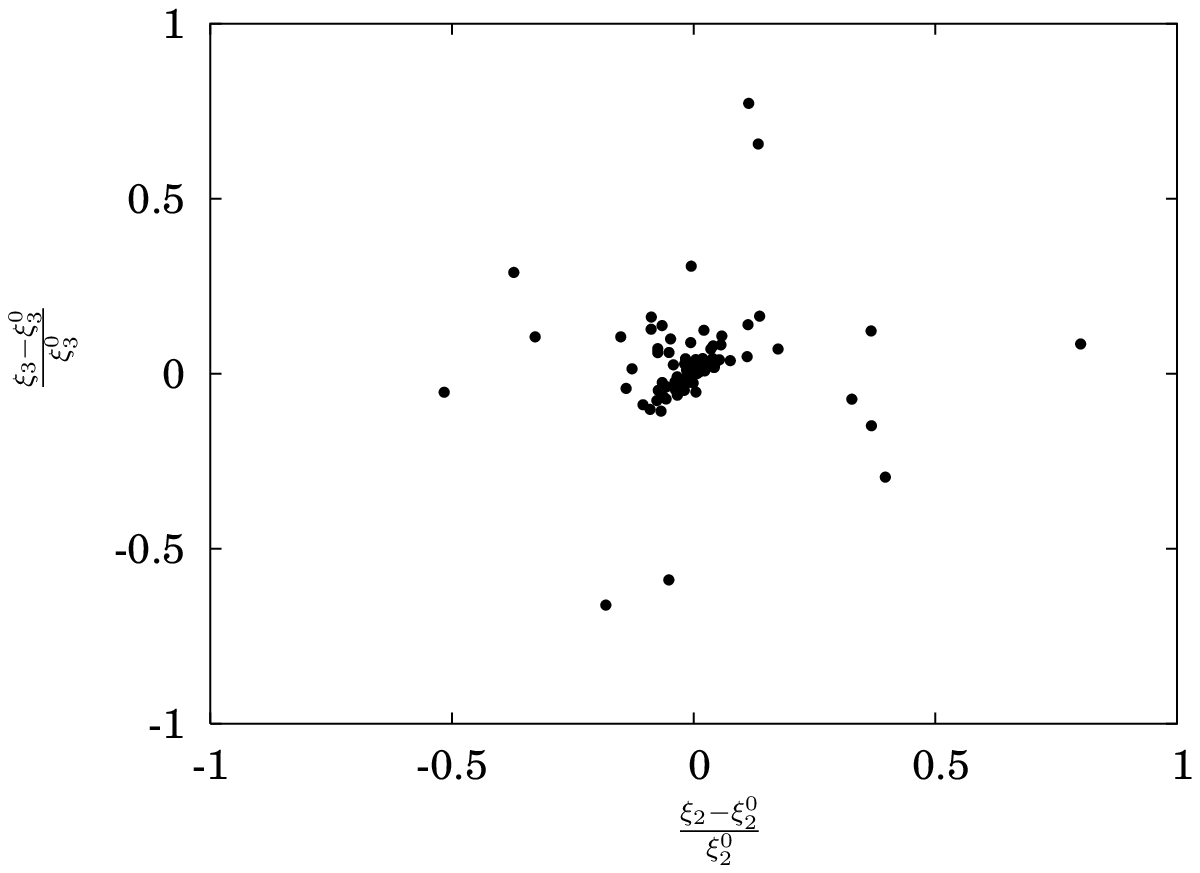 ,height=6.0cm,width=6.0cm}}
\caption{Plots for $(\xi_3-\xi_3^0)/\xi_3^0$ versus
$(\xi_2-\xi_2^0)/\xi_2^0$
showing the effect of SVEV 1-loop correction for solution
points}
\end{figure}

\begin{figure}
\epsfig{file=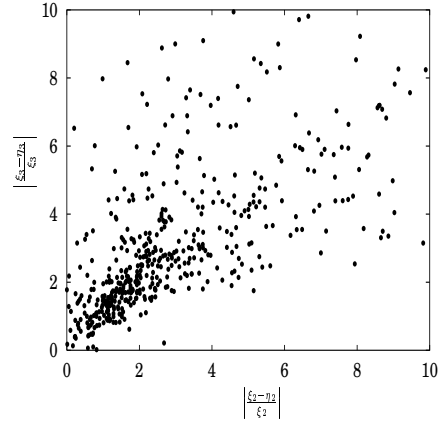, height=6.0cm,width=6.0cm} \caption{
Plots for $|(\xi_3-\eta_3)/\xi_3|$ versus $|(\xi_2-\eta_2)/\xi_2|$
showing the deviation of $\eta_i$ from $\xi_i$ direction for solution points.
}
\end{figure}

\begin{figure}
\subfigure[$\sin^2 2 \theta_{23}$ vs. $ \left| \xi_2/\xi_3 \right|
$ ] {\epsfig{file=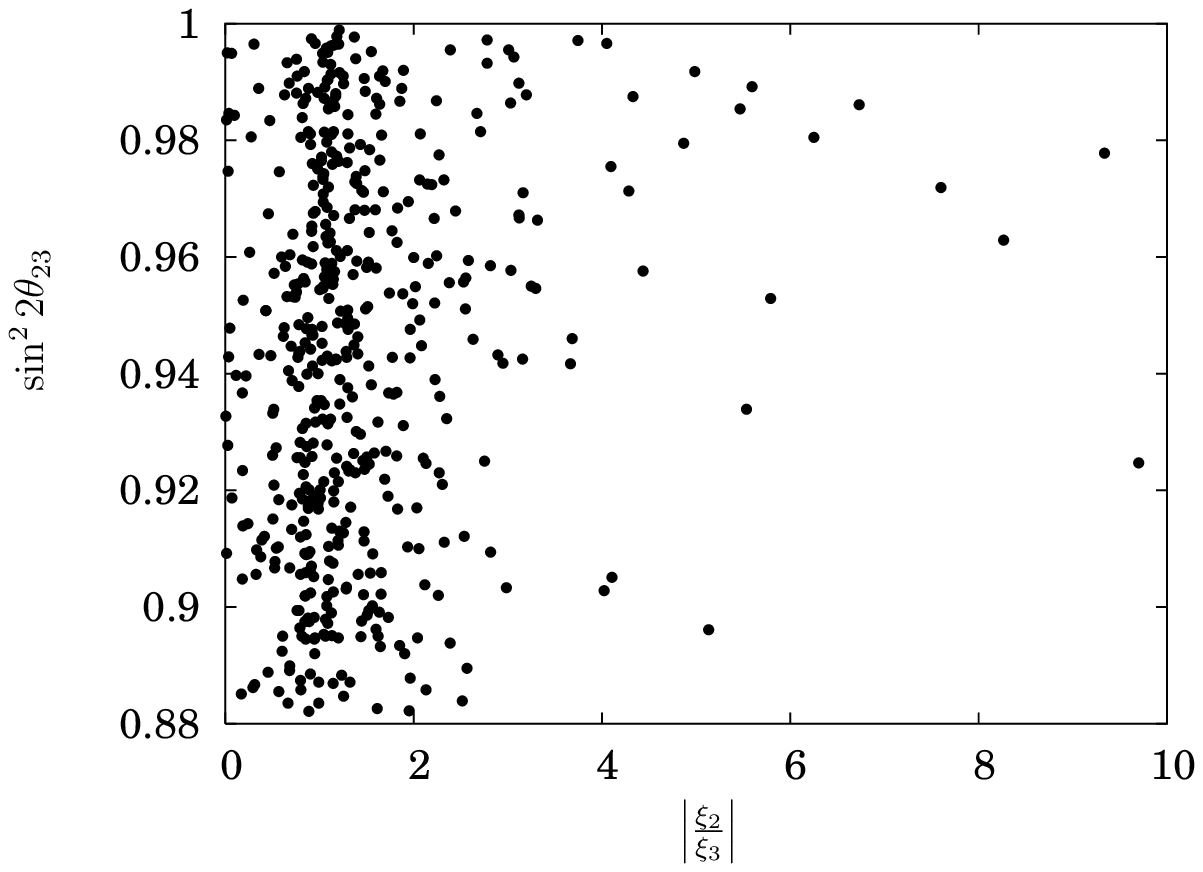
,height=6.0cm,width=6.0cm}} \subfigure[$\sin^2 2 \theta_{23}$ vs.
$ \left| \lambda_2^{'} / \lambda_3^{'} \right| $ ]
{\epsfig{file=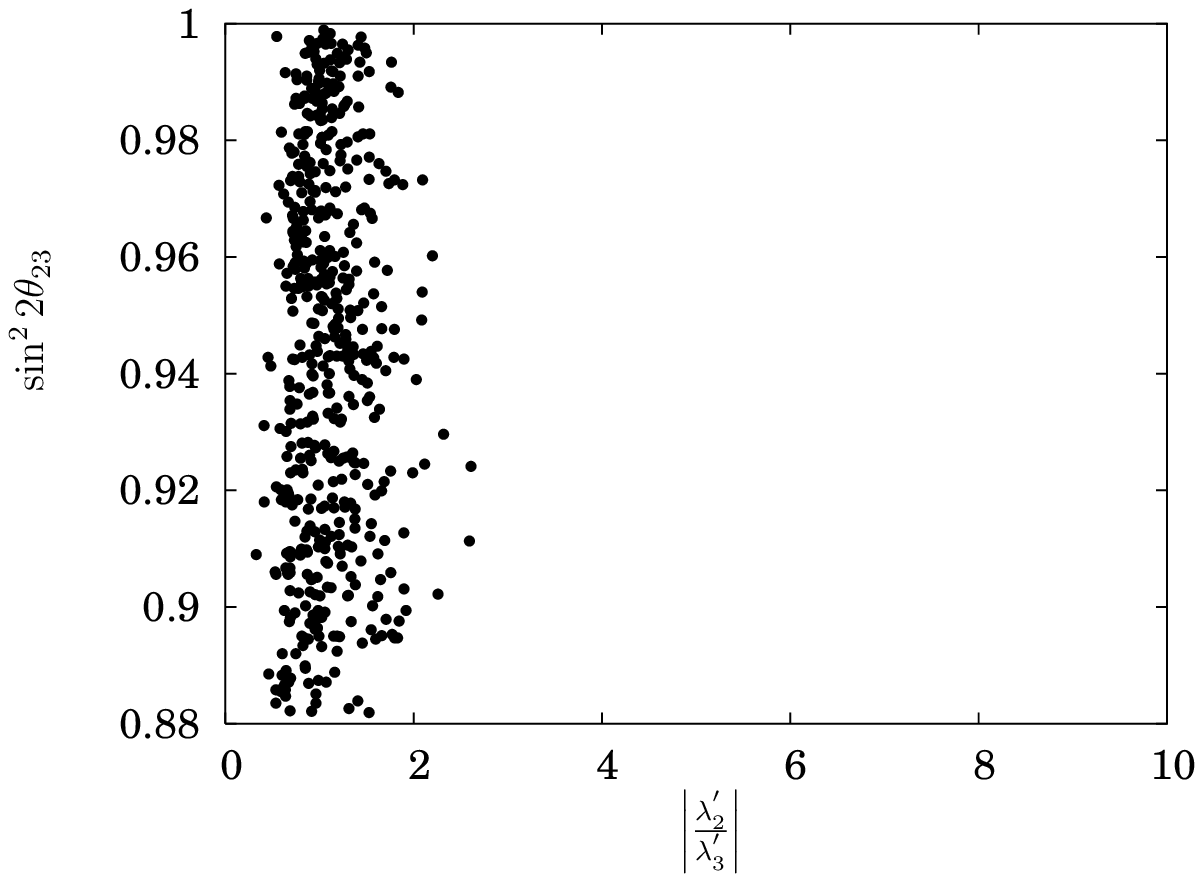 ,height=6.0cm,width=6.0cm}}
\caption{The correlation between the atmospheric neutrino mixing angle
and $|\xi_2/\xi_3|$ or $|\lambda'_2/\lambda'_3|$
 for solution points
with $\tan \beta = 3 - 15$. }
\end{figure}

\begin{figure}
\subfigure[$\sin^2 2 \theta_{23}$ vs. $ \left| \xi_3/\xi_3 \right|
$ ] {\epsfig{file=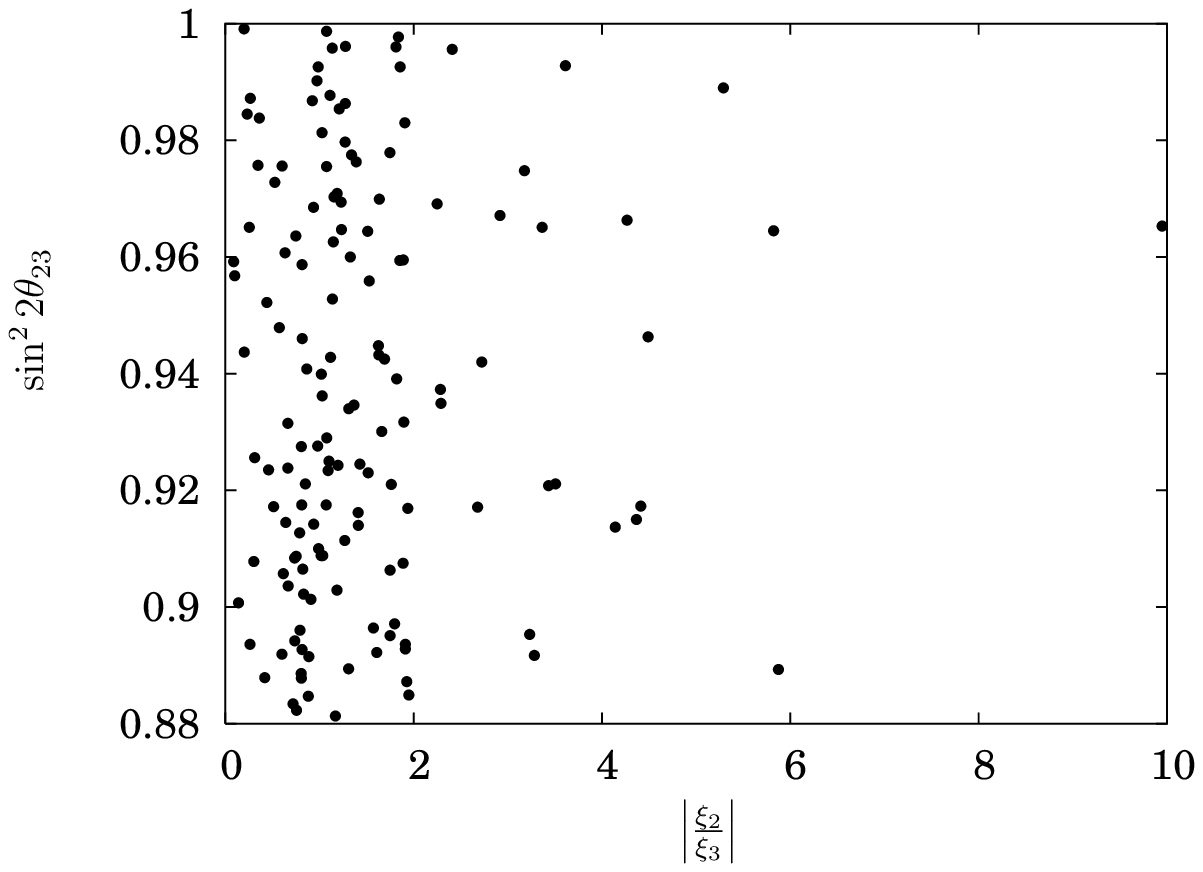,height=6.0cm,width=6.0cm}}
\subfigure[$\sin^2 2 \theta_{23}$ vs. $ \left| \lambda_2^{'} /
\lambda_3^{'} \right| $ ]
{\epsfig{file=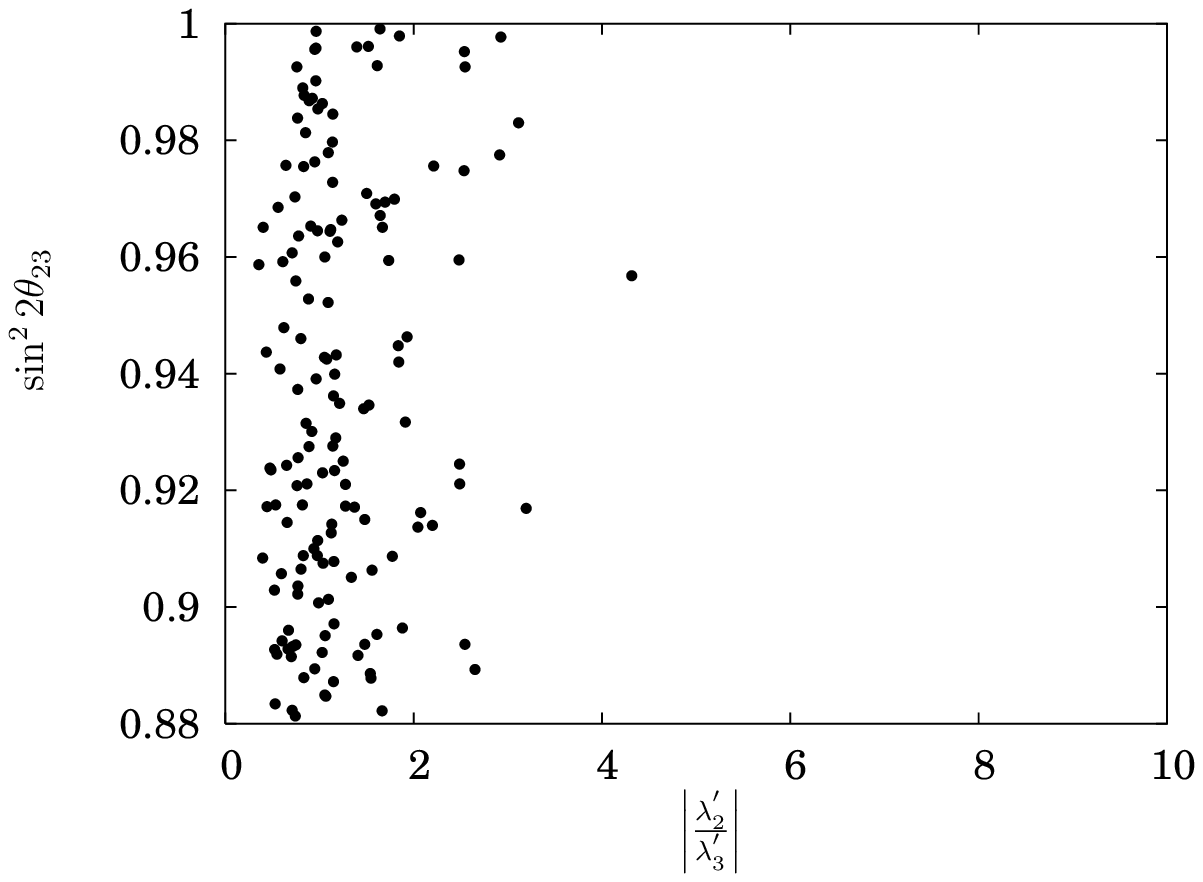,height=6.0cm,width=6.0cm}}
\caption{Same figure as above but
with $\tan \beta = 30 - 40$.}
\end{figure}

\begin{figure*}
\subfigure[$\sin^2 2 \theta_{12}$ vs. $|\lambda_1 / \lambda_2|$. ]
{\epsfig{file=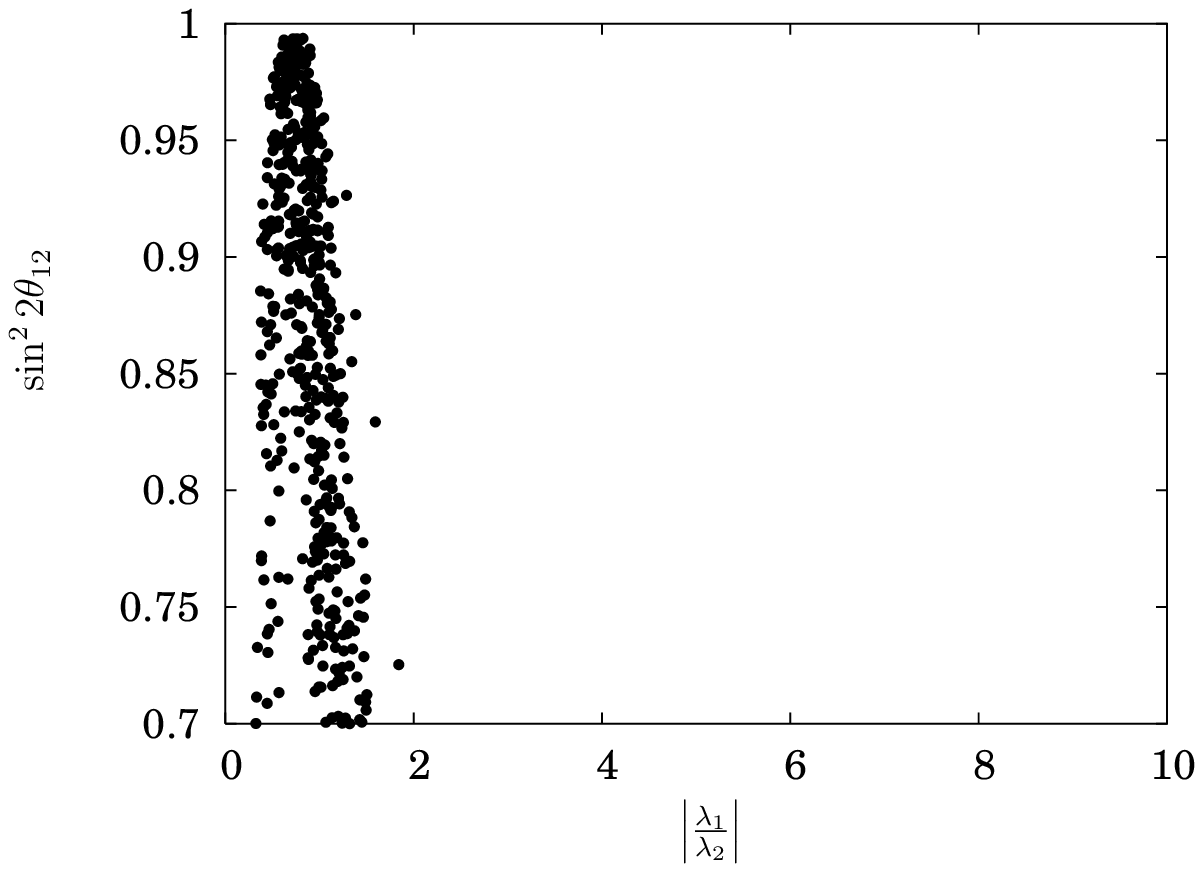,height=6.0cm,width=6.0cm}}
\subfigure[$\sin^2 2 \theta_{12}$ vs. $|\lambda_1 / \lambda_2|$. ]
{\epsfig{file=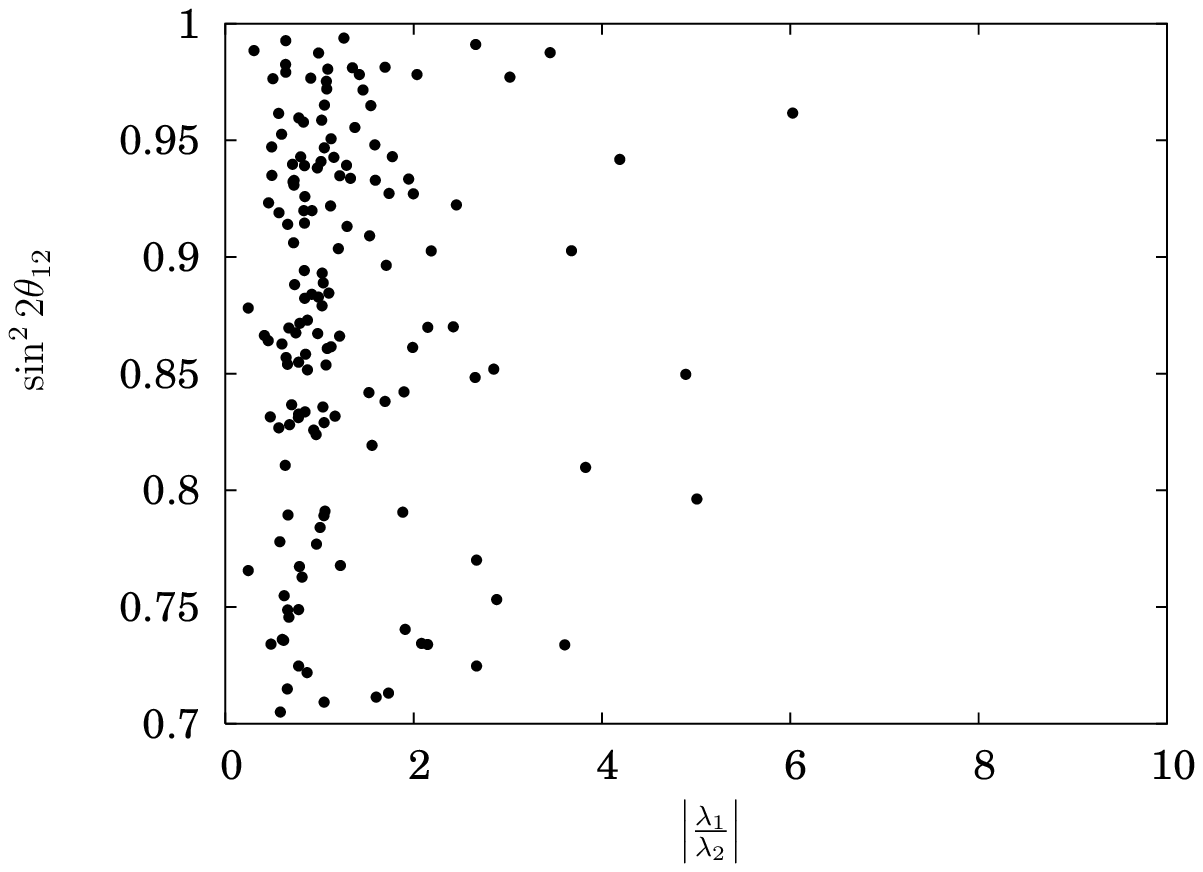,height=6.0cm,width=6.0cm}}
\caption{The correlation between solar neutrino mixing angle and
$|\lambda_1/\lambda_2|$ for solution points with
(a) $\tan \beta = 3 - 15$ and (b) $\tan \beta = 30 - 40$.}
\end{figure*}

\newpage

\begin{table}
\begin{center}
\begin{tabular}{c|ccc}
\hline
 & & Stau LSP &  \\
\hline \hline
 &  $\tan\beta=5.2$ &  $ {\rm sgn} (\mu) = -1$ &  $\mu= -801.2$ GeV
\\
 &  $ A_0 = 39.5 $GeV  &$ m_0 = 105.0  $ GeV& $M_{1/2} = 671.0$ GeV
\\
\hline
 ${\lambda}'_i$   &
$8.40\times10^{-6}$ & $-7.66\times10^{-5}$ & $-6.79\times10^{-5}$ \\
${\lambda}_i$   &
   $8.74\times10^{-5}$ & $-7.44\times10^{-5}$ & 0     \\
$\xi_i$ &
   $-9.90\times10^{-7}$& $-3.05\times10^{-6}$ & $-3.73\times10^{-6}$
\\
$\eta_i$ &
   $1.01\times10^{-6}$& $-1.31\times10^{-5}$ & $-1.19\times10^{-5}$
\\
\hline
BR  &  $e$  & $\mu$   & $\tau$   \\
\hline
$l_i \nu$ & 28.9 \%  &  21.0 \% & 50.0 \% \\
$ \bar{t}~ b$ & & $\sim$  0.07 \% & \\
\hline
       & $m_{\tilde{\tau}_1}$=278.6 GeV
     & \hfill $\Gamma=$& $2.92 \times 10^{-7}$ GeV  \\
\hline \hline
 &  \hspace{1.8cm}$\sigma_{e^+ e^- \rightarrow \tilde{\tau}_1
\tilde{\tau}_1^*}$
&$\simeq$ 1.45 $\times 10^{-2}$ (Pb), & $\sqrt{s}$ = 1 TeV  \\
\hline
\end{tabular}
\end{center}
$$ \begin{array}{l}
 (\Delta m^2_{31},~ \Delta m^2_{21})=
     (2.5\times10^{-3},~1.1\times10^{-4})~ \mbox{eV}^2  \cr
 (\sin^22\theta_{atm},~ \sin^22\theta_{sol},~ \sin^22\theta_{chooz})
 =(0.98,~0.77,~0.03) \cr
 {\rm The ~decay ~length}~L \simeq 6.74 \times 10^{-8} cm
\end{array} $$
\caption{A solution set with the stau LSP.
The values of the trilinear couplings $\tilde{\lambda}'_i$ and $\tilde{\lambda}_i$ are
taken as input parameters defined at the weak scale. }
\end{table}
\begin{table}
\begin{center}
\begin{tabular}{c|ccc}
\hline
 & &  Stau LSP &  \\
\hline \hline
 &  $\tan\beta=38.0$ &  $ {\rm sgn} (\mu) = -1$ &  $\mu= -460.1$ GeV
\\
 &  $ A_0 = 679.0 $GeV  &$ m_0 = 233.4 $ GeV& $M_{1/2} = 462.1$ GeV
\\
\hline
 ${\lambda}'_i$   &
$1.53\times10^{-6}$ & $-1.48\times10^{-5}$ & $-5.84\times10^{-6}$ \\
${\lambda}_i$   &
   $3.65\times10^{-6}$ & $5.41\times10^{-6}$ & 0     \\
$\xi_i$ &
   $-7.76\times10^{-6}$& $-2.81\times10^{-5}$ & $-2.30\times10^{-6}$
\\
$\eta_i$ &
   $8.17\times10^{-6}$& $-3.38\times10^{-5}$ & $-1.73\times10^{-5}$
\\
\hline
BR  &  $e$  & $\mu$   & $\tau$   \\
\hline
$l_i \nu$ & 11.5 \%  &  25.5 \% & 47.2 \% \\
$ \bar{t}~ b$ & & $\sim$  16.0 \% & \\
\hline
       & $m_{\tilde{\tau}_1}$=188.7 GeV
     & \hfill $\Gamma=$& $7.99 \times 10^{-10}$ GeV  \\
\hline \hline
 &  \hspace{1.8cm}$\sigma_{e^+ e^- \rightarrow \tilde{\tau}_1
\tilde{\tau}_1^*}$
&$\simeq$ 1.96 $\times 10^{-2}$ (Pb), & $\sqrt{s}$ = 1 TeV  \\
\hline
\end{tabular}
\end{center}
$$ \begin{array}{l}
 (\Delta m^2_{31},~ \Delta m^2_{21})=
     (2.5\times10^{-3},~4.6\times10^{-5})~ \mbox{eV}^2  \cr
 (\sin^22\theta_{atm},~ \sin^22\theta_{sol},~ \sin^22\theta_{chooz})
 =(0.99,~0.74,~0.005) \cr
 {\rm The ~decay ~length}~L \simeq 2.47 \times 10^{-5} cm
\end{array} $$
\caption{Same as the previous table but with large $\tan\beta$. }
\end{table}
\begin{table}
\begin{center}
\begin{tabular}{c|ccc}
\hline
 & &  Neutralino LSP &  \\
\hline \hline
 &  $\tan\beta=4.9$ &  $ {\rm sgn} (\mu) = -1$ & $\mu= -200.5$ GeV
\\
 &  $ A_0 = 38.9 $GeV  &$ m_0 = 333.7  $ GeV& $M_{1/2} = 160$ GeV
\\
\hline
 ${\lambda}'_i$   &
$-9.33\times10^{-9}$ & $-7.81\times10^{-5}$ & $-7.56\times10^{-5}$
\\
${\lambda}_i$   &
   $-5.63\times10^{-5}$ & $-7.35\times10^{-5}$ & 0     \\
$\xi_i$ &
   $-1.21\times10^{-6}$& $2.75\times10^{-7}$ & $1.83\times10^{-6}$
\\
$\eta_i$ &
   $-2.01\times10^{-6}$& $-1.17\times10^{-5}$ & $-8.74\times10^{-6}$
\\
\hline
BR  &  $e$  & $\mu$   & $\tau$   \\
\hline
$\nu jj$ &  &47.0 \%   &  \\
$ l_i^\pm jj$ &3.88 $\times 10^{-2}$ \% & 2.00 $\times 10^{-3}$\%
&
8.87 $\times 10^{-2}$\% \\
$ \nu l_i^\pm \tau^\mp$  & 9.8 \% & 16.6 \%&  26.4 \% \\
\hline
       & $m_{\tilde{\chi}^0_1}$=59.4 GeV
     & \hfill $\Gamma=$& $7.14 \times 10^{-15}$ GeV  \\
\hline \hline
 &  \hspace{1.8cm}$\sigma_{e^+ e^- \rightarrow \tilde{\chi}^0_1
\tilde{\chi}^0_1}$
&$\simeq$ 4.90 $\times 10^{-2}$ (Pb), & $\sqrt{s}$ = 1 TeV  \\
\hline
\end{tabular}
\end{center}
$$ \begin{array}{l}
 (\Delta m^2_{31},~ \Delta m^2_{21})=
     (2.5\times10^{-3},~9.6\times10^{-5})~ \mbox{eV}^2  \cr
 (\sin^22\theta_{atm},~ \sin^22\theta_{sol},~ \sin^22\theta_{chooz})
 =(0.98,~0.99,~0.008) \cr
 {\rm The ~decay ~length}~L \simeq 2.76 cm
\end{array} $$
\caption{A solution set with the neutralino LSP which is lighter than the
$W$ boson. }
\end{table}
\begin{table}
\begin{center}
\begin{tabular}{c|ccc}
\hline
 & &  Neutralino LSP &  \\
\hline \hline
 &  $\tan\beta=4.2$ &  $ {\rm sgn} (\mu) = -1$ &$\mu=-442.3$ GeV
\\
 &  $ A_0 = 98.5 $GeV  &$ m_0 = 235.9  $ GeV& $M_{1/2} = 380.9$ GeV
\\
\hline
 ${\lambda}'_i$   &
$-1.18\times10^{-8}$ & $-1.44\times10^{-4}$ & $-1.47\times10^{-4}$
\\
${\lambda}_i$   &
   $-8.57\times10^{-5}$ & $-9.71\times10^{-5}$ & 0     \\
$\xi_i$ &
   $3.24\times10^{-7}$& $-8.37\times10^{-7}$ & $-1.17\times10^{-6}$ \\
$\eta_i$ &
   $-6.72\times10^{-7}$& $-1.73\times10^{-5}$ & $-1.68\times10^{-5}$
\\
\hline
BR  &  $e$  & $\mu$   & $\tau$   \\
\hline
$\nu jj$ &  &30.1\%   &  \\
$ l_i^\pm jj$ & $3.14 \times 10^{-2}\%$ & $2.10 \times 10^{-1}\%$ &
$4.08 \times 10^{-1}\%$ \\
$ \nu l_i^\pm \tau^\mp$  & 15.1 \% & 19.4 \%& 34.4 \% \\
\hline
$l_i^{\pm}W^{\mp}$ & $2.35 \times 10^{-2} \% $ & $1.57 \times 10^{-1} \%$
& $3.04 \times 10^{-1} \%$ \\
\hline
       & $m_{\tilde{\chi}^0_1}$= 165.8 GeV
     & \hfill $\Gamma=$& $5.36\times 10^{-12}$ GeV  \\
\hline \hline
 &  \hspace{1.8cm}$\sigma_{e^+ e^- \rightarrow \tilde{\chi}^0_1
\tilde{\chi}^0_1}$
&$\simeq$ 0.12 (Pb), & $\sqrt{s}$ = 1 TeV  \\
\hline
\end{tabular}
\end{center}
$$ \begin{array}{l}
 (\Delta m^2_{31},~ \Delta m^2_{21})=
     (2.5\times10^{-3},~6.5\times10^{-5})~ \mbox{eV}^2  \cr
 (\sin^22\theta_{atm},~ \sin^22\theta_{sol},~ \sin^22\theta_{chooz})
 =(0.99,~0.96,~0.001) \cr
 {\rm The ~decay ~length}~L \simeq 3.67 \times 10^{-3} cm
\end{array} $$
\caption{A solution set with the neutralino LSP which is heavier
than the $W$ boson but lighter than the top quark }
\end{table}

\begin{table}
\begin{center}
\begin{tabular}{c|ccc}
\hline
& &  Neutralino LSP &  \\
\hline \hline
&  $\tan\beta=7.6$ &  $ {\rm sgn} (\mu) = -1$ &$\mu=-832.3$ GeV
\\
&  $ A_0 = 48.2 $GeV  &$ m_0 = 255.4 $ GeV& $M_{1/2} = 715.3$ GeV
\\
\hline
${\lambda}'_i$   &
$-5.48\times10^{-6}$ & $-4.99\times10^{-5}$ & $-7.26\times10^{-5}$
\\
${\lambda}_i$   &
$-5.00\times10^{-5}$ & $-6.22\times10^{-5}$ & 0     \\
$\xi_i$ &
$2.12\times10^{-6}$& $7.66\times10^{-6}$ & $8.53\times10^{-6}$ \\
$\eta_i$ &
$-3.41\times10^{-7}$& $-5.84\times10^{-6}$ & $-9.05\times10^{-6}$
\\
\hline
BR  &  $e$  & $\mu$   & $\tau$   \\
\hline
$\nu jj$ &  &11.6\%   &  \\
$ l_i^\pm jj$ & 1.82$\times 10^{-1}$ \% & 2.37 \% & 2.95 $
$ \% \\
$ l_i^\pm (tb)$ & $ 5.81\times 10^{-2}$\% & $ 1.65 $\%
& $3.13 $\% \\
$ \nu l_i^\pm \tau^\mp$  & $14.8$ \% & $22.9$ \%
& $37.0$ \% \\
\hline
$l_i^{\pm} W^{\mp} $ & $1.35 \times 10^{-1} \% $ & 1.77 \% &
2.19 \% \\
\hline
& $m_{\tilde{\chi}^0_1}$=315.5 GeV
& \hfill $\Gamma=$& $2.14\times 10^{-11}$ GeV  \\
\hline \hline
&  \hspace{1.8cm}$\sigma_{e^+ e^- \rightarrow \tilde{\chi}^0_1
\tilde{\chi}^0_1}$
&$\simeq$ $0.14$ (Pb), & $\sqrt{s}$ = 1 TeV  \\
\hline
\end{tabular}
\end{center}
$$ \begin{array}{l}
(\Delta m^2_{31},~ \Delta m^2_{21})=
(2.5\times10^{-3},~2.2\times10^{-5})~ \mbox{eV}^2  \cr
(\sin^22\theta_{atm},~ \sin^22\theta_{sol},~ \sin^22\theta_{chooz})
                   =(0.91,~0.97,~0.05) \cr
{\rm The ~decay ~length}~L \simeq 9.23 \times 10^{-4} cm
\end{array} $$
\caption{ A solution set with the neutralino LSP which is heavier than
the top quark  but lighter than two top quarks.}
 \end{table}

%

\begin{table}
\begin{center}
\begin{tabular}{c|ccc}
\hline
& &  Neutralino LSP &  \\
\hline \hline
&  $\tan\beta=9.7$ &  $ {\rm sgn} (\mu) = -1$ &$\mu=-1045.8$ GeV
\\
&  $ A_0 = 526.5 $GeV  &$ m_0 = 308.5 $ GeV& $M_{1/2} = 947.6$ GeV
\\
\hline
${\lambda}'_i$   &
$9.00\times10^{-6}$ & $6.54\times10^{-5}$ & $-7.61\times10^{-5}$
\\
${\lambda}_i$   &
$6.40\times10^{-5}$ & $7.51\times10^{-5}$ & 0     \\
$\xi_i$ &
$-3.49\times10^{-6}$& $6.97\times10^{-6}$ & $-3.82\times10^{-6}$ \\
$\eta_i$ &
$-3.56\times10^{-6}$& $1.96\times10^{-5}$ & $-2.12\times10^{-5}$
\\
\hline
BR  &  $e$  & $\mu$   & $\tau$   \\
\hline
$\nu t \bar{t}$ &  &97.8\%   &  \\
$ l_i^\pm jj$ & 4.6$\times 10^{-3}$ \% & 1.8$\times 10^{-2}$\% & 5.5$\times
10^{-3}$\% \\
$ l_i^\pm (tb)$ & $ 1.6\times 10^{-3}$\% & $ 3.3 \times 10^{-2}$\%
& $4.3 \times 10^{-2}$\% \\
$ \nu l_i^\pm \tau^\mp$  & $4.3\times 10^{-1}$ \% & $5.9\times 10^{-1}$ \%
& $1.02$ \% \\
\hline
$l_i^{\pm} W^{\mp}$ & $3.39 \times 10^{-3} \%$ & $1.35 \times 10^{-2}$ \% &
$4.05 \times 10^{-3}$ \%  \\
\hline
& $m_{\tilde{\chi}^0_1}$=417.0 GeV
& \hfill $\Gamma=$& $2.04\times 10^{-9}$ GeV  \\
\hline \hline
&  \hspace{1.8cm}$\sigma_{e^+ e^- \rightarrow \tilde{\chi}^0_1
\tilde{\chi}^0_1}$
&$\simeq$ $1.04\times 10^{-1}$ (Pb), & $\sqrt{s}$ = 1 TeV  \\
\hline
\end{tabular}
\end{center}
$$ \begin{array}{l}
(\Delta m^2_{31},~ \Delta m^2_{21})=
(3.6\times10^{-3},~3.2\times10^{-5})~ \mbox{eV}^2  \cr
(\sin^22\theta_{atm},~ \sin^22\theta_{sol},~ \sin^22\theta_{chooz})
                  =(0.96,~0.98,~0.03) \cr
{\rm The ~decay ~length}~L \simeq 9.66 \times 10^{-6} cm
\end{array} $$
\caption{A solution set with the neutralino LSP which is heavier than
two top quarks.}
\end{table}

\end{document}